\documentclass[%
reprint,
superscriptaddress,
showpacs,preprintnumbers,
 amsmath,amssymb,
 aps,
]{revtex4-1}

\usepackage{graphicx}
\usepackage{dcolumn}
\usepackage{bm}

\usepackage{references}
\usepackage{hyperref} 

\usepackage{equationarray}
\usepackage[normalem]{ulem}
\usepackage[usenames]{xcolor}
\usepackage{booktabs}

\newcommand{\be}{\begin{equation}}
\newcommand{\ee}{\end{equation}}
\newcommand{\ba}{\begin{eqnarray}}
\newcommand{\ea}{\end{eqnarray}}
\newcommand{\bd}{\begin{displaymath}}
\newcommand{\ed}{\end{displaymath}}

\newcommand{\vv}[1]{\textcolor{blue}{ \bf VK:  #1}}
\usepackage{ulem} 

\newcommand{\prak}[1]{\textcolor{cyan}{\bf MP: #1}}

\begin{document}

\preprint{APS/123-QED}

\title{Shear viscosity of hadrons with K-matrix cross sections}

\author{Anton Wiranata }
\email{awiranata@lbl.gov}
\affiliation{Institute of Particle Physics and Key Laboratory of Quark and Lepton Physics (MOE), Central China Normal University, Wuhan 430079, China}
\affiliation{Lawrence Berkeley National Laboratory, Nuclear Science Division, MS 70R0319, Berkeley, CA 94720, USA}

\author{Volker Koch}
\email{vkoch@lbl.gov}
\affiliation{Lawrence Berkeley National Laboratory, Nuclear Science Division, MS 70R0319, Berkeley, CA 94720, USA}

\author{Madappa Prakash}
\email{prakash@phy.ohiou.edu}
\affiliation{Department of Physics and Astronomy, Ohio University, Athens, OH, 45701}

\author{Xin Nian Wang}
\email{xnwang@lbl.gov}
\affiliation{Institute of Particle Physics and Key Laboratory of Quark and Lepton Physics (MOE), Central China Normal University, Wuhan 430079, China}
\affiliation{Lawrence Berkeley National Laboratory, Nuclear Science Division, MS 70R0319, Berkeley, CA 94720, USA}

\date{\today}

\begin{abstract}
Shear viscosity $\eta$ and entropy density $s$ of a hadronic resonance gas are calculated 
using the Chapman-Enskog and virial expansion methods
using the $K$-matrix parameterization of hadronic cross sections which 
preserves the unitarity of the $T$-matrix. In the $\pi-K-N-\eta$ mixture considered, a total of 57 resonances up to 2 GeV were included.
Comparisons are also made to results with other hadronic cross sections such as 
the Breit-Wigner (BW) and, where available, experimental phase shift  
parameterizations. 
Hadronic interactions forming resonances are shown
to decrease the shear viscosity  and increase the entropy density   
leading to a substantial reduction of $\eta/s$ as the QCD phase transition temperature is approached.

\end{abstract}

\maketitle


\section{Introduction}

Experimental data from high-energy nucleus-nucleus collisions at the Relativistic 
Heavy-Ion Collider (RHIC) \cite{star:10,star:12a,star:12b,star:12c,Phenix:10a,Phenix:10b,Phenix:11a,Phenix:12a} and the Large Hadron Collider (LHC)
\cite{Alice:10a, Alice:11, Alice:11a, Alice:13, Alice:13a, Alice:13b,
  Atlas:12, CMS:12a, CMS:12b, CMS:13} 
indicate that a significant amount of collective flow is being developed 
during the partonic or quark-gluon
plasma stage of the evolving system. The collective flow is manifested
in  particle spectra which are blue shifted and exhibit an anisotropy
with respect to the event plane. 
The anisotropy in the spectra, which is  often referred
  to as anisotropic flow, is due to 
collective expansion which converts the geometric anisotropy of the initial energy density distribution
into anisotropic distribution in azimuthal angles of the 
detected hadrons. The observed anisotropic 
flow is typically characterized in terms of Fourier components, $v_n$. For
non-central collisions, the second moment, $v_2$ dominates reflecting
the shape of the overlapping nuclei at finite impact parameters. The
higher Fourier components, on the other hand are due to event-by-event
fluctuations of the initial geometry \cite{Alver-Roland:10}. These
higher moments have   
been studied within event-by-event ideal and viscous hydrodynamical model 
calculations \cite{Pang:12, Hannah:11, Bleicher:11, Jan:10, Shen:11,
  Hirano:11}.  
with the finding that 
the they are mainly determined by 
fluctuations in the initial conditions, 
and the ratio of the shear viscosity $\eta$ and entropy density $s$. Comparisons between experimental data and 
hydrodynamic calculations 
thus have the promise to 
provide increasingly stringent constraints on the 
$\eta/s$ of the dense matter and the initial conditions \cite{Huchao:09,RR:07,Schanke:11}.	

First principle calculations of the shear viscosity of strongly interacting matter from QCD 
over a large range of temperature have so far been elusive. 
However, model calculations of strongly interacting matter
in different ranges of temperature
can provide an interesting picture of the temperature dependence 
of the shear viscosity to entropy density ratio, $\eta/s$
\cite{Kapusta:06,CK:10}.
Based on experimental $\pi-\pi$ shifts,  calculations \cite{PPVW:93}  give an $\eta/s$ ratio for 
a pion gas that decreases with temperature as $\sim 1/T^4$ \citep{Chenwang:07b}, 
whereas perturbative QCD 
calculations \cite{Yaffe:00, Yaffe:03, Xu:08, Xu:11, Chenwang:10, Chenwang:11} 
predict  a smaller $\eta/s$ that increases logarithmically 
with temperature in a quark-gluon plasma. 
This ratio,  believed to be bound by 
a limit of $1/(4\pi)$ (in units of $\hbar/k_B$)
from AdS/CFT calculations \cite{Son:05},  
likely reaches a minimum around the QCD phase transition temperature.

Lacking firm theoretical guidance on the temperature 
dependence of the $\eta/s$ ratio, most hydrodynamic model studies have assumed a 
constant $\eta/s$, whereas 
some studies have incorporated hadronic cascade models (termed afterburners) 
\cite{Huchao:11,Jeon:13} 
following hydrodynamic evolution in order to incorporate the effect of 
a large shear viscosity in the hadronic phase. 
A systematic study of shear viscosity to entropy 
density ratio in the hadronic phase is therefore important 
to assess the effects of transport 
properties throughout the evolution, that is, from of 
the quark-gluon plasma stage to the phase transition stage and thereafter the hadronic stage. 

In this paper, we present our study of the shear viscosity to entropy density ratio in 
a hadronic resonance gas with 
57 resonances (with masses up to 2 GeV taken from 
the Particle Data book) formed by interactions among the 
components of a $\pi-K-N-\eta$ mixture.  
Our calculations are for a system with zero net baryon and strangeness numbers.
Resonant interactions, including the widths of the various resonances,  are 
incorporated consistently in 
calculations of both the shear viscosity and the entropy 
density. We 
employ the $K$-matrix parametrization of the hadronic cross sections which 
accommodates multiple resonances and preserves the unitarity of the $T$-matrix in all channels. 
We show how the inclusion of 
multiple resonances 
in a multi-component mixture 
decreases 
the shear viscosity to entropy density ratio $\eta/s$ with increasing number of components in the mixture as 
the temperature approaches that of the QCD phase transition.

This paper is organized as follows.
Our calculation of shear viscosity, carried out numerically within the Chapman-Enskog 
approximation, 
is reviewed in Sec. II.  In Sec. III, we 
examine different 
parameterizations of 
differential cross sections that have been used 
in the calculation of transport 
cross sections needed for the calculation of shear viscosity.  
The thermodynamics of interacting hadrons 
using the virial expansion 
approach is discussed in Sec. IV. 
Analytical and numerical 
results are presented in Sec V for a single component gas of pions. 
Section VI contains results for the multi-component $\pi-K-N-\eta$ mixture. We 
summarize and conclude with an outlook in Sec. VI.
The Appendices contain relevant details of the K-Matrix, Breit-Wigner and phase shift parametrizations of cross sections.

\section{Shear Viscosity}
\label{Formalism}

We employ the Chapman-Enskog approach, generalized to include relativistic kinematics, for the 
calculation of shear viscosity. This approach enables us to consider a mixture in which particles with different  masses
give rise to multiple massive resonances through hadronic interactions. 
In the case of a single-component system, we follow closely the formalism developed in 
Refs.~\cite{Polak:73, Kox:75}  considering only number conserving elastic processes. 
In the case of  binary and multicomponent  mixtures, we adapt the formalism detailed in Refs.~\cite{Kox:75, Leeuwen:75}
to treat the hadronic gas composed of a large number of resonances. In the context of hadronic interactions, the procedure to calculate 
shear viscosity has also been described in detail in Refs.~\cite{PPVW:93,Wiranata:12}.
We present only the final results in this section and refer the reader to earlier works for details.

\subsection{Single Component System}
\label{singlecomponent}
Here we summarize the working formulae relevant for a
single component gas consisting of  hadrons with mass $m$. To first order, the shear viscosity is given by
\begin{eqnarray}
\eta_s = \frac{1}{10}\,T\,\frac{\gamma_0^2}{c_{00}}\,,
\quad \gamma_0 = -10\,\hat{h}\,,\,\,\,\hat{h} = K_3(z)/K_2(z)\,,
\label{shearone}
\end{eqnarray}
where $z = m/T$ is the relativity parameter, $\hat{h}$ is the reduced enthalpy per particle, and $K_\nu(z)$ is the modified
Bessel function of order $\nu$. The quantity $c_{00}$ is given by 
\begin{eqnarray}
c_{00}(z) = 16\,(w_2^{(2)}(z) -w_1^{(2)}(z)/z + w_0^{(2)}(z)/(3z^2))\,,
\label{c00}
\end{eqnarray}
where $w_i^{(s)}(z)$ are the so-called relativistic omega integrals:  
\begin{eqnarray}	
w_{i}^{(s)}(z) &=& \frac{2\pi z^3}{K_2(z)^2}\int_{0}^{\infty} d\psi
\sinh^7  \psi \cosh^i\psi K_j(2z\cosh\psi)\nonumber \\
&& \times\,\int_{0}^{\pi} d\Theta \sin \Theta \,\sigma_{00}(\psi,\Theta)(1-\cos^s\Theta)~.
\label{relomega}
\end{eqnarray}
The differential cross section for interaction between two identical particles is given 
by  $\sigma_{00}(\psi, \Theta)$, $j = 5/3 + 1/2(-1)^i$. Hyperbolic functions of the quantity $\psi$
characterize the relative momentum and invariant center-of-mass energy of the two colliding particles:
\begin{eqnarray}
\sinh \psi = \frac{g}{m}\,,\,\,\, \cosh \psi = \frac{P}{2m}\,,
\end{eqnarray}
where $g = \sqrt{(p_1 - p_2)^2}/2$, $P = \sqrt{(p_1+p_2)^2}$, $p_1$ 
and $p_2$ are the initial four-momenta of
the two colliding hadrons. 

The Chapman-Enskog approach also allows us to improve upon 
the first order result in Eq.~(\ref{shearone}). Expressions for higher orders can be found 
in Refs.~\cite{Polak:73, Kox:75}. Results obtained using higher order approximations
in the case 
of a pion gas with experimental cross sections have been presented earlier in Ref.~\cite{PPVW:93}
showing rapid convergence with respect to the order of the approximation.    

\subsection{Multi-Component System}
\subsubsection{Binary Mixture}

To first order, 
the shear viscosity in the case of a binary mixture is given by
\begin{equation}
\eta_s = \frac{1}{10}T\left[ \frac{\gamma_{2}^2~c_{11} 
+ \gamma_1^2~c_{22} - 2~\gamma_{1}~\gamma_2~ c_{12}}{\left( c_{11}c_{22}-c_{12}^2\right) }\right],
\label{shear1st}
\end{equation}
where $\gamma_k = -10\,\tilde{x}_k\hat{h}_k$ ($k$=1,2) and 
$\hat{h}_k = K_3(z_k)/K_2(z_k)$  with $z_k = m_k/T$ being the reduced enthalpy 
per particle for particle type $k=(1,2)$ with mass $m_k$. The coefficients
$\tilde{x}_k = \rho_k/\rho$ are related to the mass density 
$\rho_k$ (mass times the number density) of particle type 
$k$ and $\rho = \sum_k \rho_k$ is the total mass density. 
The coefficients $c_{kl}$ are given by
\begin{eqnarray}
c_{kk} &=& c_{00}(z_k) + \tilde{c}_{kk}(kl);
\label{singleterm} \\
c_{kl} &=& \tilde{c}_{kl}(kl)\,,\quad \textnormal{for}\quad k \neq l \,,
\label{diffterm}
\end{eqnarray}
%
where the coefficients $\tilde{c}_{kl}(kl)=\tilde{c}_{lk}(lk)$ are used to denote contributions to the shear viscosity from interactions 
between different particle species ( $1-2$ for a binary mixture), while $c_{00}(z_k)$ accounts for contributions from
interaction between two identical particles of type $k$. The corresponding coefficients $\tilde{c}_{kl}(kl)$ are by given by
\begin{eqnarray}
\tilde{c}_{12}&=&\frac{32\rho^2\tilde{x}_1^2\tilde{x}_2^2}{3M_{12}^2n^2x_1x_2}
\left[ -10z_1z_2\zeta^{-1}_{12}Z_{12}^{-1}w_{1211}^{(1)}(\sigma_{12}) \right.\nonumber\\
&&\left. - 10z_1z_2\zeta^{-1}_{12}Z_{12}^{-2}w_{1311}^{(1)}(\sigma_{12})
+3w_{2100}^{(2)}(\sigma_{12}) \right.\nonumber\\
&&\left. -3Z_{12}^{-1}w_{2200}^{(2)}(\sigma_{12}) + Z_{12}^{-2} w_{2300}^{(2)}(\sigma_{12})\right]~,
\label{cmin1p1}\\
\tilde{c}_{11} &=&\frac{32\rho^2\tilde{x}_1^2\tilde{x}_2^2}{3M_{12}^2n^2x_1x_2}
\left[ 10z_1^2\zeta^{-1}_{12}Z_{12}^{-1}w_{1220}^{(1)}(\sigma_{12})  \right.\nonumber\\
&&\left. + 
10z_1^2\zeta^{-1}_{12}Z_{12}^{-2}w_{1320}^{(1)}(\sigma_{12})
+3w_{2100}^{(2)}(\sigma_{12}) \right.\nonumber\\
&&\left.-3Z_{12}^{-1}w_{2200}^{(2)}(\sigma_{12}) + Z_{12}^{-2} w_{2300}^{(2)}(\sigma_{12})\right]~,
\label{cp1p1}\\
\tilde{c}_{22} &=&\frac{32\rho^2\tilde{x}_1^2\tilde{x}_2^2}{3M_{12}^2n^2x_1x_2}
\left[ 10z_2^2\zeta^{-1}_{12}Z_{12}^{-1}w_{1202}^{(1)}(\sigma_{12}) \right.\nonumber\\
&&\left.+
10z_2^2\zeta^{-1}_{12}Z_{12}^{-2}w_{1302}^{(1)}(\sigma_{12})
+ 3w_{2100}^{(2)}(\sigma_{12}) \right.\nonumber\\
&&\left.-3Z_{12}^{-1}w_{2200}^{(2)}(\sigma_{12}) + Z_{12}^{-2} w_{2300}^{(2)}(\sigma_{12})\right]\,,
\label{cmin1min1}
\end{eqnarray}
where $x_k=n_k/n$, $n_k$ is the particle number density of particle
type $k$ and $n=\sum_k n_k$ is the total particle number density. 
We have denoted the summed mass by $M_{kl}= m_k +m_l$ and the reduced mass by $\mu_{kl} = m_km_l/M_{kl}$ for
two nonidentical particles ($k\neq l$). Additionally,  $\zeta_{kl} = 2\mu_{kl}/T$ and  $Z_{kl} = M_{kl}/2T$. 

The relativistic omega integrals for a binary mixture are given by
\begin{eqnarray}
 w_{rtuv}^{(s)}(\sigma_{kl}) &=& \frac{\pi\,\mu_{kl}\,}{4\,T\,K_2(z_k)\,K_2(z_l)}\,\int_0^\infty\,d\Psi_{kl}\, \sinh^3\Psi_{kl}\nonumber \\
& \times& \left(\frac{g_{kl}^2}{2\mu_{kl}T} \right)^r \left(\frac{M_{kl}\,}{P_{kl}} \right)^t \left(\cosh\,\psi_k \right)^u \left(\cosh\,\psi_l \right)^v \nonumber \\
&\times& K_n \left(\frac{P_{kl}}{T} \right)\,\int_0^\pi\,d\Theta_{kl}\,\sin \Theta_{kl}\,\sigma_{kl}\left(\Psi_{kl},\Theta_{kl} \right) \nonumber \\
&\times& \left( 1 - \cos^s\Theta_{kl} \right)\,,
\label{omegabinary}
\end{eqnarray}
where
\begin{eqnarray}
 P_{kl}^2 &=& m_k^2 + m_l^2 + 2m_k m_l \cosh\,\Psi_{kl} , \\
 g_{kl}P_{kl} &=& m_km_l  \sinh\,\Psi_{kl} , \quad \Psi_{kl} \equiv \psi_k + \psi_l \\
 \cosh\,\psi_k &=& \frac{1}{P_{kl}} \left(m_k + m_l \cosh\, \Psi_{kl}\right) , \\
  \cosh\,\psi_l &=& \frac{1}{P_{kl}} \left(m_l + m_k \cosh\, \Psi_{kl}\right)\, ,
\end{eqnarray}
with  $P_{kl}=\sqrt{(p_k+p_l)^2}$ and $g_{kl}=\sqrt{(p_k-p_l)^2}/2$ characterizing the invariant 
center-of-mass energy and relative momentum of two particles with initial four-momentum $p_k$ and $p_l$.
In the case of $m_k = m_l = m_i$,   $\Psi_{ii} = 2\psi_i$, $P_{ii} = 2m_i\,\cosh \psi_i$
and the relativistic omega integral for binary mixture  reduces to Eq.~(\ref{relomega}).

\subsubsection{Tertiary and Higher Component Mixtures  }

In a mixture containing $N$ components $(k,l=1,2,...,N)$,  
the first order coefficient of shear viscosity 
within the
Chapman-Enskog approximation is given by 
\begin{eqnarray}
 \eta_s = \frac{1}{10}\,\rho^2T^3\,\sum_{k=1}^N\,\sum_{l=1}^N\,\,c_{k}\, 
 c_{l}\,c_{kl}\,,\,\,\nonumber \\
\end{eqnarray}
where  $c_{k}$ are coefficients of the orthogonal Laguere polynomials
used as an ansatz function in solving for the coefficient of shear viscosity (for  
details, see Refs.  \cite{LPG:73, Leeuwen:75}).
The additional sum rule required to solve for $c_{k}$ in terms of 
$c_{kl}$ and $\gamma_{k}$ is given by
\begin{eqnarray}
 \sum_{l=1}^N\,\left( \frac{\rho\,T}{n} \right)^2\,c_{l}\,
c_{kl} = \frac{\rho\, T}{n^2}\,\gamma_{k}\,\,,
\end{eqnarray}
where $k,l = 1, 2, \cdots, N$ (the number of particle species).

As for the single component gas and for the binary mixture, the coefficients
$c_{k}$ are linear combinations of $\gamma_k$ and $c_{kr}$. Thermodynamic
variables are hidden in the coefficients $\gamma$. Interactions  among
particles in the mixture given in terms of  differential cross-sections reside
in the omega integrals and the coefficients $c_{kr}$ are linear combinations of
the omega integrals \cite{Leeuwen:75}.
Explicitly, the required coefficients can be written as 
    \begin{eqnarray}
     c_{kk} &=&  16\,\tilde{x}_k^2\,\left( w_2^{(2)}(\sigma_{kk}) - \frac{1}{z}\,w_1^{(2)}(\sigma_{kk})
 + \frac{1}{3z}w_0^{(2)}(\sigma_{kk}) \right) \nonumber \\
 &&+  \sum_{l\neq k }^N\frac{32\rho^2\tilde{x}_k^2\tilde{x}_l^2}{3M_{kl}^2n^2x_kx_l}
 \left[ 10z^2_k\zeta_{kl}^{-1}Z_{kl}^{-1}w_{1220}^{(1)}(\sigma_{kl})\right. \nonumber \\
  && \left. +10z_k^2\zeta_{kl}^{-1}Z_{kl}^{-1}w_{1320}^{(1)}(\sigma_{kl})
+3\,w_{2100}^{(2)}(\sigma_{kl})\right.\nonumber\\
&&\left.-3Z_{kl}^{-1}w_{2200}^{(2)}(\sigma_{kl}) + Z_{kl}^{-1} w_{2300}^{(2)}(\sigma_{kl})\right]~
\label{singlemixture}\\
\tilde{c}_{kl}&=&\frac{32\rho^2\tilde{x}_k^2\tilde{x}_l^2}{3M_{kl}^2n^2x_kx_l}
\left[ -10z_kz_l\zeta^{-1}_{kl}Z_{kl}^{-1}w_{1211}^{(1)}(\sigma_{kl}) \right.\nonumber\\
&&\left. - 10z_kz_l\zeta^{-1}_{kl}Z_{kl}^{-2}w_{1311}^{(1)}(\sigma_{kl})
+3w_{2100}^{(2)}(\sigma_{kl}) \right.\nonumber\\
&&\left. -3Z_{kl}^{-1}w_{2200}^{(2)}(\sigma_{kl}) + Z_{kl}^{-2} w_{2300}^{(2)}(\sigma_{kl})\right]~
\,(k\neq l)~.
\label{multimixture}
    \end{eqnarray}

The first of the above equations, 
Eq. (\ref{singlemixture}),  tells us
how similar particles interact  in the presence of other types of particles, whereas the second equation, 
Eq. (\ref{multimixture}), tells us how dissimilar particles interact in the mixture.

\section{Cross Section Parametrizations}
\label{crosssection}

The magnitude of shear viscosity is strongly determined by the strength of interactions between the constituent particles in a system. 
As evident from Eqs. (\ref{shearone})-(\ref{relomega}), the shear viscosity is 
inversely proportional to the differential cross section of the interacting particles. 
Large cross sections, characteristic of a  strongly interacting system, naturally lead to small viscosities.  
In this work, we focus on shear viscosities in the hadronic phase of the strongly interacting matter 
that is created at RHIC/LHC.  For all but the lightest particles, first principle calculations of hadronic interactions, particularly those involving
massive resonances do not exist. One therefore often uses empirical parameterizations of these hadronic cross sections. 
To assess the impact of different parameterizations on the shear viscosity,  we examine three forms of parameterizations:   
(i) cross sections obtained directly from the phase shifts \cite{Bertsch:88,PPVW:93}, 
(ii) the Breit-Wigner parametrization for Ref.~\cite{Bass:99}, 
and (iii) cross sections from the $K-$matrix~\cite{Chung:95} parametrization.

\subsection{Cross Sections from  $\pi - \pi$ phase shifts}
\begin{figure}
	\includegraphics[width=9cm]{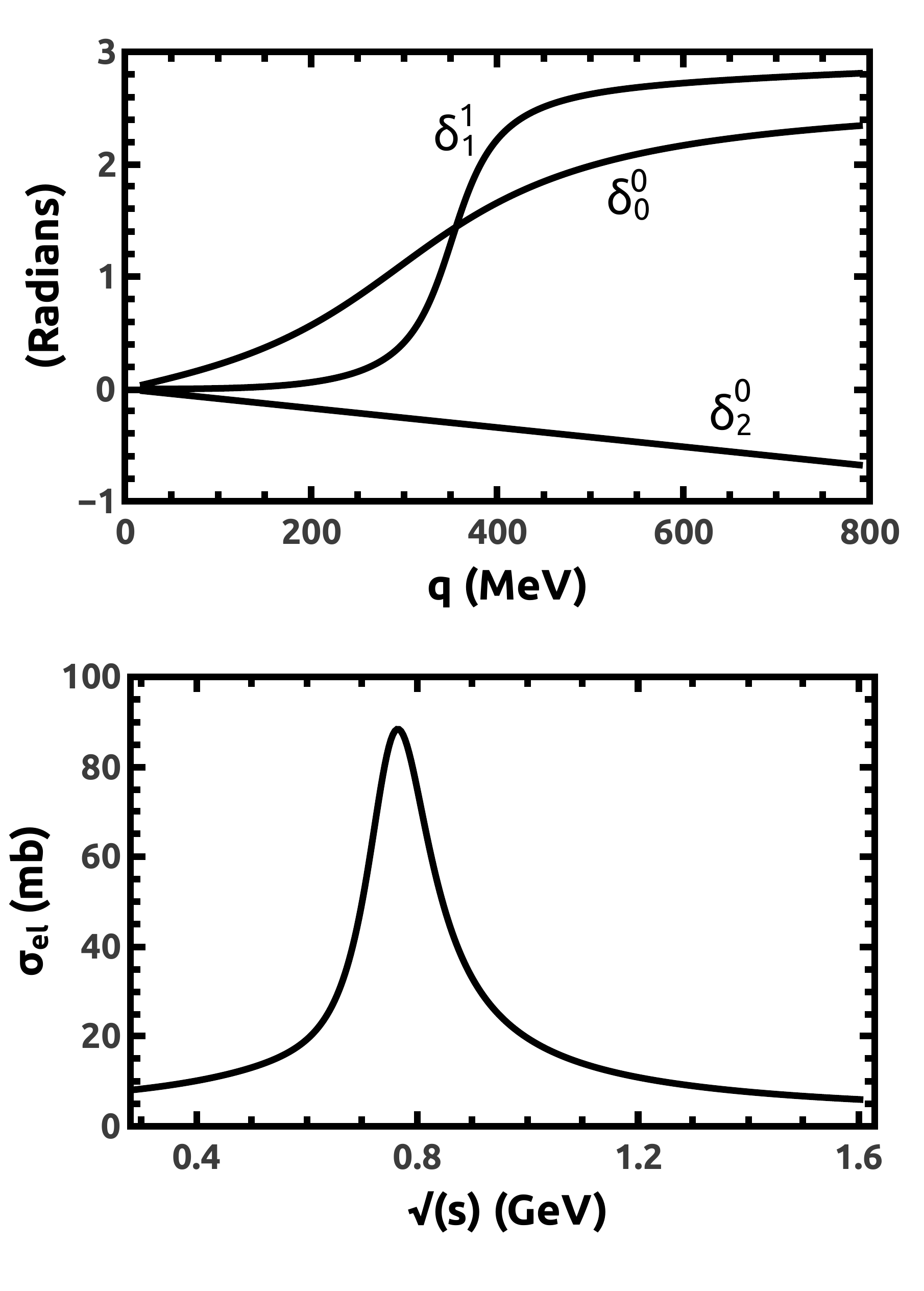}
	\caption{Upper panel: $\pi-\pi$ phase shifts versus center-of-mass momentum from 
	Eqs. \ref{pssigma}, \ref{repulsivephaseshift} and \ref{rhophaseshift}.
	Lower panel: Total cross section versus center-of-mass energy from
	Eq. (\ref{phaseshiftcross}). }
\label{phasecross}
\end{figure}

A pion gas represents a good test case in which the role of resonances on the viscosity is exemplified. 
Experimentally, $\pi-\pi$ shifts are available at least in three channels. The impact of additional channels for which phase shifts are not 
available will be considered in a subsequent section.
For practical and illustrative purposes, 
we follow the parametrization employed by Bertsch et al. \cite{Bertsch:88}, which was used in the calculation of the shear viscosity of
a pion gas in \cite{PPVW:93}.  In terms of phase shifts, the differential cross section can be parameterized by
\begin{eqnarray}
\sigma(\sqrt{s},\theta) &=& \frac{4}{q^2} \sum_{l+I={\rm even}} {}\frac{(2I+1)(2l+1)}{\sum_I (2I+1)} \nonumber \\
&&\times P_l^2(\cos\,\theta)\,\sin^2\delta_l^I(\sqrt{s})\,, \nonumber \\
\end{eqnarray}
where $l$ is the orbital angular momentum, $I$ is the isospin and
 summation is over both $l$ and $I$ for  $l+I$ being even numbers. 
The phase shift $\delta_0^0$, corresponding to the 
$\sigma-$resonance,  is well fit by 
\begin{eqnarray}
 \delta_0^0 &=& \frac{\pi}{2} + \arctan\left( \frac{\sqrt{s}-m_{\sigma}}{\Gamma_{\sigma}/2} \right)\,,\,\,\,
  \Gamma_{\sigma} = 2.06\,q\,,
  \label{pssigma}
\end{eqnarray}
where $m_\sigma = 5.8 m_\pi$ is the mass of the $\sigma-$resonance, $m_\pi$ is the pion mass,
$\sqrt{s}$ the center-of-mass energy and $q = \sqrt{s - 4 m_{\pi}^2}/2$ the center-of-mass momentum of two colliding pions.
The phase shift $\delta_0^2$ corresponds to the repulsive channel and is given by
\begin{eqnarray}
 \delta_0^2 &=& - 0.12\,\frac{q}{m_{\pi}}\,.
 \label{repulsivephaseshift}
\end{eqnarray}
The phase shift $\delta_1^1$ corresponds to the  $\rho-$resonance and is expressed by 
\begin{eqnarray}
 \delta_1^1 &=& \frac{\pi}{2} + \arctan \left(\frac{E-m_{\rho}}{\Gamma_{\rho}/2}  \right)\,,\label{rhophaseshift} \\
 \Gamma_{\rho} &=& 0.095\,q\,\left[\frac{q/m_{\pi}}{(1+(q/m_\rho)^2)}  \right]^2\,,
\end{eqnarray}
where $m_\rho=5.53  m_\pi$ is the mass of the $\rho-$resonance. 
Including contributions from $l=0,1,2$,  the differential cross section is then
\begin{eqnarray}
\sigma(\sqrt{s},\theta) &=& \frac{4}{q^2} \left( \frac{1}{9}\sin^2\delta_0^0 +
\frac{5}{9}\sin^2\delta_0^2 + 3\,\sin^2\delta_1^1\cos^2\theta \right)\,.\nonumber \\
\label{phaseshiftcross}
\end{eqnarray}
These three phase shifts
and the total cross section are shown in Fig. \ref{phasecross} as a function of center of 
mass momentum and energy, respectively.  

\subsection{The Breit-Wigner Parametrization}
The interaction or $T$-matrix for hadronic interaction through a single resonance 
$(a+b \rightarrow R \rightarrow a+b)$
can also be parameterized in the Breit-Wigner form \cite{Yao:06} as
\begin{equation}
T=\frac{m_R\Gamma_{R\rightarrow ab}(\sqrt{s})}{(m_R^2-s) - im_R
  \Gamma_R^\textnormal{tot} (\sqrt{s})}\, .
\end{equation}
where $ \Gamma_R^\textnormal{tot}=\sum_{c,d} \Gamma_{R\rightarrow cd}$ is the
total width and $\Gamma_{R\rightarrow ab}$ is the partial width for the 
channel $R\rightarrow ab$ of the resonance $R$, respectively.
The differential cross section for such an interaction is then
\begin{eqnarray}
 \sigma(\sqrt{s},\theta) &=& \frac{C(I,l)}{q^2_{ab}}\,\frac{m_R^2\Gamma_{R\rightarrow ab}^2}{(m_R^2-s)^2 +m_R^2 {\Gamma_R^\textnormal{tot}}^2}
 P_l(\cos\,\theta) \nonumber \\
 &\approx&  \frac{C(I,l)}{q^2_{ab}}\,\frac{\Gamma_{R\rightarrow ab}^2}{4(m_R-\sqrt{s})^2 + {\Gamma_R^\textnormal{tot}}^2}
 P_l(\cos\,\theta) \,, \nonumber\\
 \label{bwcross}
\end{eqnarray}
where $C(I,l)$ is the symmetry factor which contains the spin-isopin multiplicities for the corresponding
resonance and $q_{ab}$ the center-of-mass momentum. The general cross
section for the reaction $a+b \rightarrow c+d$ is then obtained by
integrating 
over the polar angle and
summing over all relevant resonances
\begin{eqnarray}
\sigma(\sqrt{s})_{a+b \rightarrow c+d} &=& \sum_{R} \frac{2S_R+1}{(2S_a+1)(2S_b+1)}\,\frac{\pi}{q^2_{ab}} \nonumber \\
& \times& \frac{\Gamma_{R\rightarrow ab}\Gamma_{R\rightarrow cd}}{(m_R-\sqrt{s})^2 + {\Gamma_R^\textnormal{tot}}^2/4}\,,
\label{urqmdtotcross}
\end{eqnarray}
where $I$ denotes the isospin, $I_3$ is the third component of the isospin and
$S$ is the spin for hadrons and resonances. The coefficients with angular brackets are 
the Clebsch-Gordon coefficients for the isospin. The center-of-mass
momentum of the incoming particles is
\begin{eqnarray}
	q_{ab}(\sqrt{s}) = \frac{1}{2\sqrt{s}}\sqrt{(s-(m_a+m_b)^2)(s-(m_a-m_b)^2)}\,, \nonumber \\
\end{eqnarray}
and the energy dependence of the width in a given channel is typically
given by
\begin{eqnarray}
\Gamma_{R\rightarrow ab}(\sqrt{s}) &=& \Gamma^0_{R\rightarrow ab} \frac{m_R}{\sqrt{s}}
\left( \frac{q_{ab}(\sqrt{s})}{q_{ab}(m_R)} \right)^{2l+1}\nonumber \\
&\times&\frac{1.2}{1+0.2\left(\frac{q_{ab}(\sqrt{s})}{q_{ab}(m_R)} \right)^{2l}}\,,
\end{eqnarray}
where $m_R$ is the mass of the resonance, $\Gamma_{R\rightarrow ab}^0$ is the width for the
channel $R\rightarrow ab$ at the pole, 
$l$ is the orbital angular momentum of the exit (decay) 
channel. Values for the resonance masses and their decay widths at 
the pole can be found in the Review of Particle Physics \cite{Yao:06}.
The last term in the above equation is related to the 
Blatt-Weisskopf $B$-factor which can be found in \cite{Weisskopf:89}.

\subsection{The $K-$Matrix Parametrization}
\label{Kmatrixsection}
The $K$-matrix formalism \cite{Wigner:46,Wigner:47,Chung:95} preserves 
unitarity of the $T$-matrix for processes of the type
$ab \rightarrow cd$. In this section, we provide a brief summary following  
closely the exposition in Ref.~\cite{Chung:95}.
The differential cross section for $ ab \rightarrow cd$ is given in terms of 
the invariant amplitude $\cal{M}$ (or the scattering amplitude $f$)  as
\begin{eqnarray}
\sigma(\sqrt{s},\theta) = \frac{1}{(8\pi)^2 s}\,\left(\frac{q_{cd}}{q_{ab}} \right)
 |\mathcal{M} |^2 = |f (\sqrt{s},\theta)|^2\,,
 \label{diffcross}
\end{eqnarray}
where $q_{ab}(q_{cd})$ is the breakup momentum in the initial (final) state; 
$\theta$ is the usual polar angle in spherical coordinate 
and $\sqrt{s}$ is the center-of-mass energy. The scattering amplitude $f(\theta)$
can be expressed as
\begin{eqnarray}
 f(\sqrt{s}, \theta) = \frac{1}{q_{ab}} \sum_l \left(2l + 1 \right) \, T^l(s) \,P_l(\cos\theta)\,,
\end{eqnarray}
in terms of the interaction matrix $T^l(s)$. The Legendre polynomials  $P_l(\cos\theta)$ account for the
angular momentum dependence of the exit channel. 

In general, the $T$ matrix can be defined from the overlap matrix between the
initial and final state of the collision:
\begin{eqnarray}
 S_{ab\rightarrow cd} = \langle cd | S | ab \rangle\,, \quad S = I+ 2i\,T\,,
\end{eqnarray}
where $S$ is the scattering operator (matrix), $I$ is the identity operator accounting for no interaction.
Based on the unitarity of the $S-$matrix,
\begin{eqnarray}
 S\,S^\dag &=& S^\dag \, S = I\,,
\end{eqnarray}
one can arrive at  the relation
\begin{eqnarray}
  \left(T^{-1} + iI \right)^\dag &=& T^{-1} + iI\, .
\end{eqnarray}
Therefore, one can define a Hermitian $K-$matrix through
\begin{eqnarray}
 K^{-1} = T^{-1} + iI \,,\quad K = K^\dag\,.
\end{eqnarray}
Time reversal symmetry of $S$ and $T$ also leads to $K$ being symmetric. Therefore, the $K$-matrix can be chosen
to be real and symmetric. One can rewrite the $T$ matrix in terms of the $K-$matrix as
\begin{eqnarray}
 {\rm Re}\, T &=& (I+K^2)^{-1}K = K(I+K^2)^{-1} \nonumber \\
 {\rm Im}\, T &=& (I+K^2)^{-1}K^2 =K^2 (I+K^2)^{-1}\,.
 \label{tmatrixink}
\end{eqnarray}

In the $K$-matrix formalism, resonances appear as a sum of poles in the $K-$matrix:
\begin{eqnarray}
 K_{ab\rightarrow cd} = \sum_R \frac{g_{R\rightarrow ab }(\sqrt{s})g_{R\rightarrow cd}(\sqrt{s})}{m_R^2 - s}\,,
 \label{kresonances}
\end{eqnarray}
where the sum on $R$ goes over the number of resonances with masses $M_R$.
The decay couplings are given by
\begin{eqnarray}
 g_{R\rightarrow ab }^2(\sqrt{s}) = m_R \Gamma_{R\rightarrow ab}(\sqrt{s})\,,
\end{eqnarray}
where the partial decay widths are given by 
\begin{eqnarray}
 \Gamma_{R\rightarrow ab} (\sqrt{s}) = \Gamma_{R\rightarrow ab}^0 \frac{m_R}{\sqrt{s}} 
 \frac{q_{ab}}{q_{ab0}}  [B^l(q_{ab},q_{ab0})]^2\,
\end{eqnarray}
with $q_{ab0}=q_{ab}(m_R)$ being the breakup momentum at $\sqrt{s} =
M_R$ and $\Gamma^0$ the width at the pole, as defined previously.
The $B^l(q,q_0)$ is  the usual Blatt-Weisskopf barrier factors which can be written in terms of the
breakup momentum in channel $R\rightarrow ab $ and the resonance breakup momentum $q_{ab}$ for the
orbital angular momentum $l$:
\begin{eqnarray}
  B_{R\rightarrow ab}^l(q_{ab},q_{ab0}) = \frac{F_l(q_{ab})}{F_l(q_{ab0})}\,.
\end{eqnarray}
The list of $F_l(q)$ for $l=0$ through 4 reads as
\begin{eqnarray}
 F_0(q) &=& 1\,, \quad F_l(1) = 1 \nonumber \\
 F_1(q) &=& \sqrt{\frac{2z}{z+1}}\,, \quad  F_2(q) = \sqrt{\frac{13z^2}{(z-3)^2 +9z}}\nonumber \\
 F_3(q) &=& \sqrt{\frac{277z^3}{z(z-5)^2 +9(2z-5)}}\,\,\,\,\textnormal{and} \nonumber \\
 F_4(q) &=& \sqrt{\frac{12746z^4}{(z^2-45z+105)^2 +25z(2z-21)^2}}\,,
\end{eqnarray}
where $z = (q/q_R)^2$ and $q_R = 0.1973$ GeV/c.  For interaction with a single resonance in the
intermediate state, one can verify that the Breit-Wigner and $K$-matrix parameterizations are identical.
We will explore their differences for the case of multiple, and especially, overlapping resonances later
in this section.

\subsection{Comparisons}
In this subsection, we first compare results of total cross sections from 
three different parameterizations as described in the previous subsection for  the 
$\pi \pi \rightarrow \rho \rightarrow \pi \pi$ channel. 
Thereafter, we include other resonances in $\pi \pi$ reactions in our comparisons. 
Finally, a discussion in the case of overlapping resonances is provided.

\subsubsection{Single resonance} 
\begin{figure}
\includegraphics[width=9.5cm]{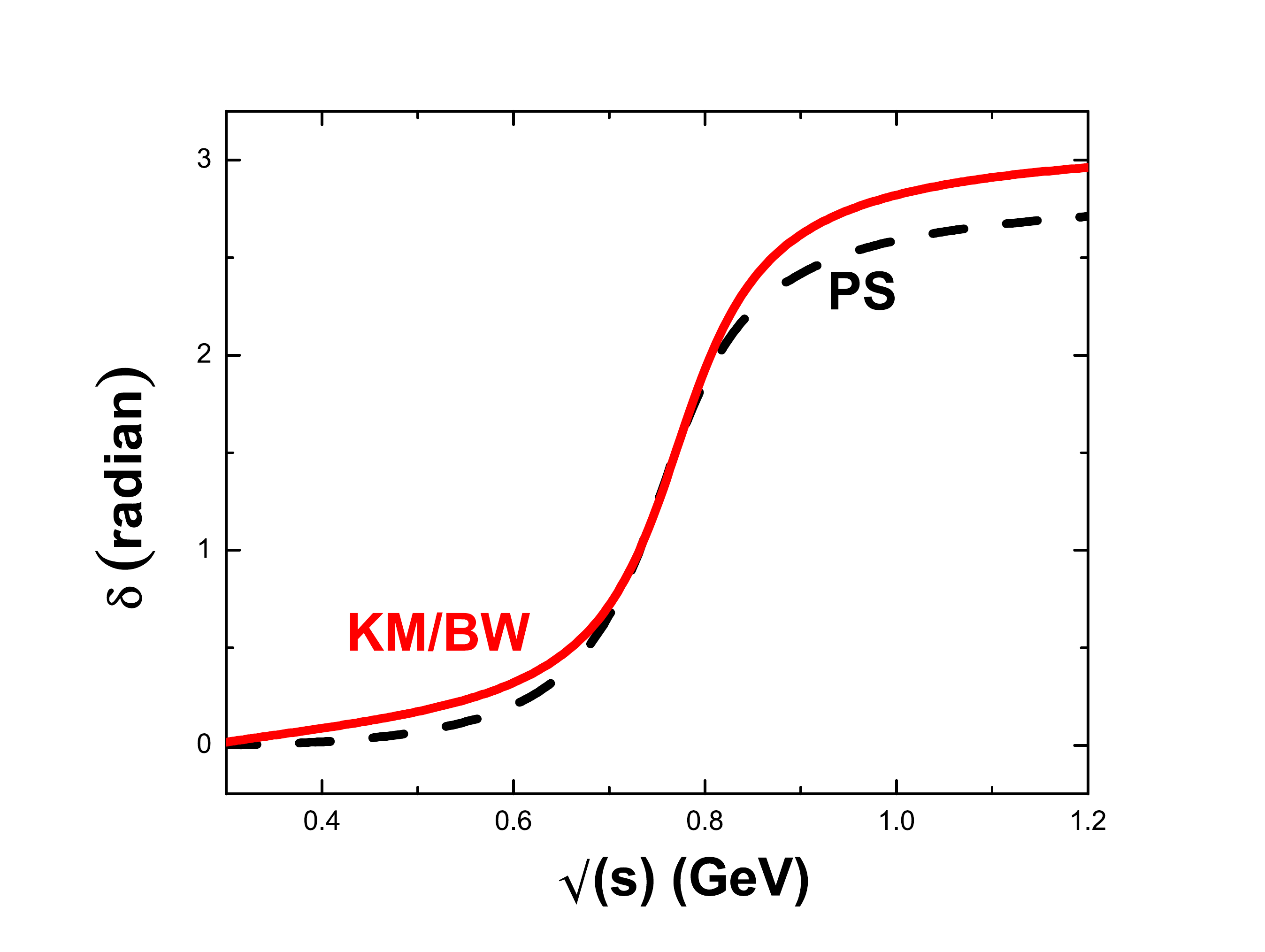}
\caption{Phase shifts for $\pi\pi \rightarrow \rho \rightarrow \pi \pi$ 
from experiment as it is parameterized in Eq. \ref{rhophaseshift}
and from $K-$matrix from Eq. (\ref{phaseandTandK}). }
\label{totalphaseshift}
\end{figure}

For interaction through a single resonance, the Breit-Wigner (BW) and $K$-matrix (KM) parameterizations
are nearly identical. We will compare them to the phase  shift (PS) parameterizations here. Similar
to the partial wave decomposition of the $T$-matrix,
\begin{eqnarray}
 T &=& e^{i\delta_l} \sin\delta_l \,,
 \label{phaseandT}
\end{eqnarray}
one can relate the phase shift in a single resonance channel to the $K$-matrix,
\begin{eqnarray}
 K &=& \frac{m_\rho\,\Gamma_{\rho\rightarrow \pi\pi} (\sqrt{s})}{m_\rho^2 - s} = \tan \delta_l\,
 \label{phaseandTandK}
\end{eqnarray}
for the process  $\pi\,\pi \rightarrow \rho \rightarrow \pi \pi$.
In Fig. \ref{totalphaseshift}, we compare the phase shift from BW/KM and PS parameterizations
from Eq. (\ref{rhophaseshift}).  We used $m_\rho = 0.77$ GeV and 
the $\Gamma_{\rho\rightarrow \pi\pi}^0 = 0.15$ GeV. The symmetry factor $C(I,l)$ is the same 
for all formalisms. 
While there is good agreement between the different parameterizations near the peak of the 
resonance, some differences exist at threshold and high energies
because of the 
differences in the parameterizations of the widths between the PS and
KM/BW approaches. 

As discussed in Appendices A and B, all three parametrizations lead to
identical cross sections if the same energy dependence of the decay width
($\Gamma_\rho$) is employed. 

\begin{figure}
\includegraphics[width=9.5cm]{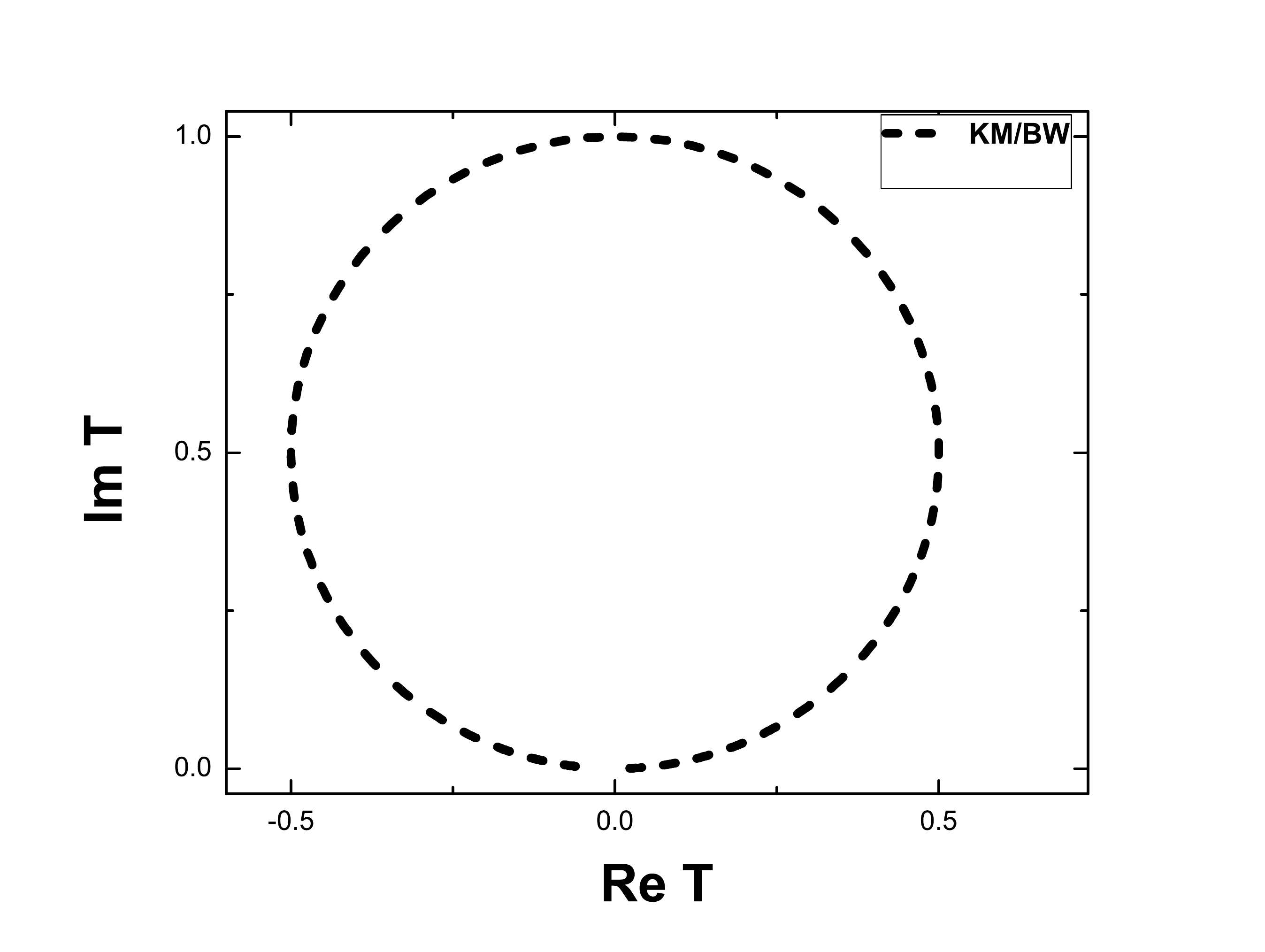}
\caption{The argand plot of the $T-$matrix for the $\pi \pi \rightarrow \rho \rightarrow \pi \pi$
calculated from K-matrix and Breit-Wigner formalisms. }
\label{Tmatrixrhoonly}
\end{figure}

As we can see from the Argand diagram in Fig. \ref{Tmatrixrhoonly}, 
both the KM/BW parametreizations for a single resonance maintain the unitarity of the $T-$matrix.

\begin{figure}
\includegraphics[width=9cm]{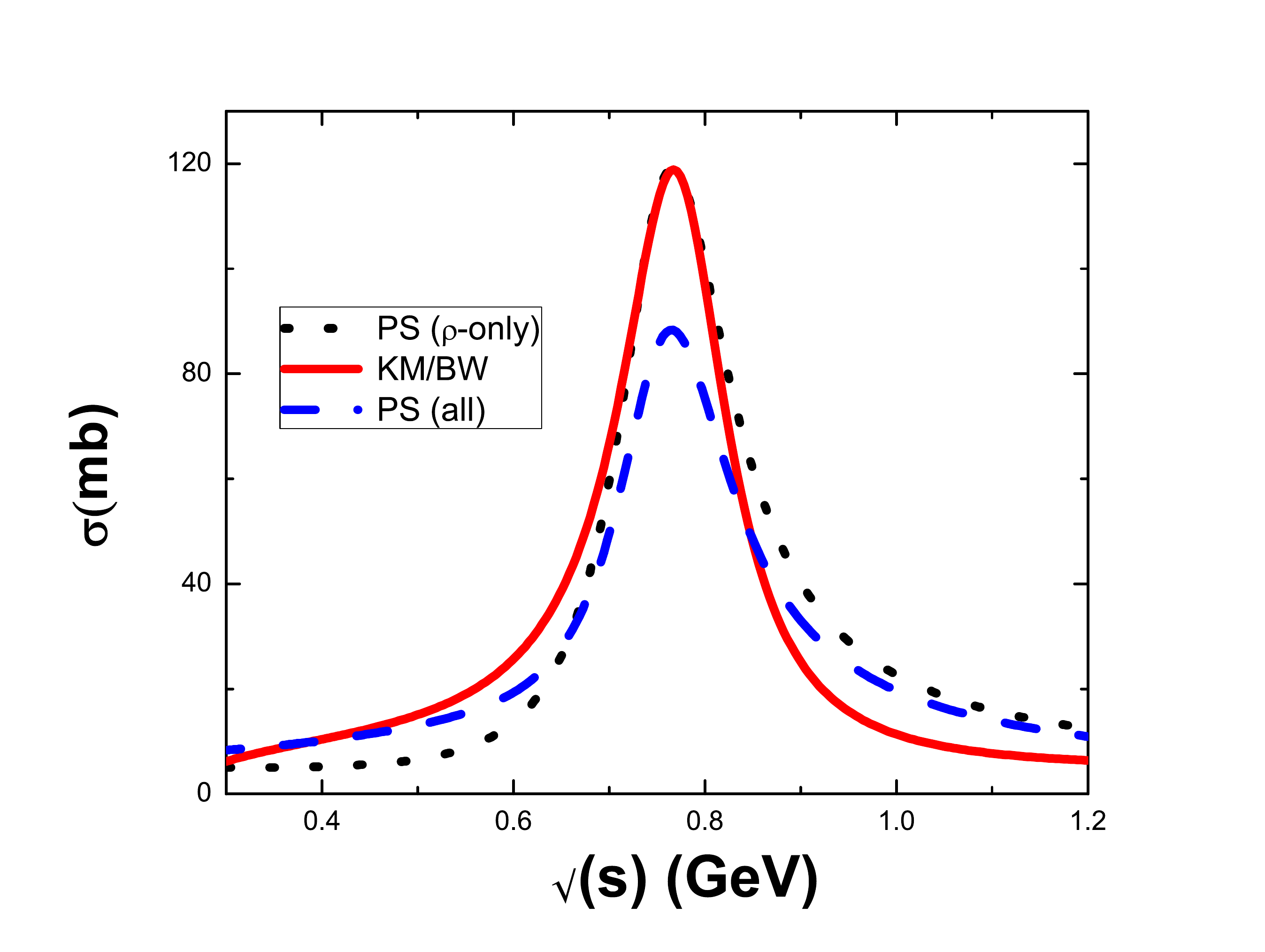}
\caption{Total cross sections from phase shift fits, the Breit-Wigner/Urqmd and K-matrix formalisms. }
\label{totalcrosssection}
\end{figure}

In Fig. \ref{totalcrosssection}, we show how a fit to the experimental cross sections 
that includes all three phase shifts compares with
the KM/BW cross sections when only the $\rho-$ resonance is included. In the results shown,  
the fit to the experimental cross sections is isospin averaged. 
By using the appropriate  symmetry factor, one is able to
get the same result as that in the BW cross section.

In Fig. \ref{totalcrosssection}, we also compare the total cross sections from  BW/KM parameterizations (solid curve),
the PS parameterization with the $\rho$ resonance only (dot-dashed), and the PS parameterization with the $\sigma$ and $\rho$ resonances
together with the repulsive channel (dashed). While cross sections from BW/KM and PS for a single resonance are very similar
around the resonance peak, there are significant differences near the threshold and at higher energies. The inclusion
of the $\sigma$ resonance and the repulsive channel in the PS parameterization significantly reduces the cross section at
the resonance peak.

\subsubsection{Separated and Overlapping Resonances}

\begin{figure}
\includegraphics[width=9cm]{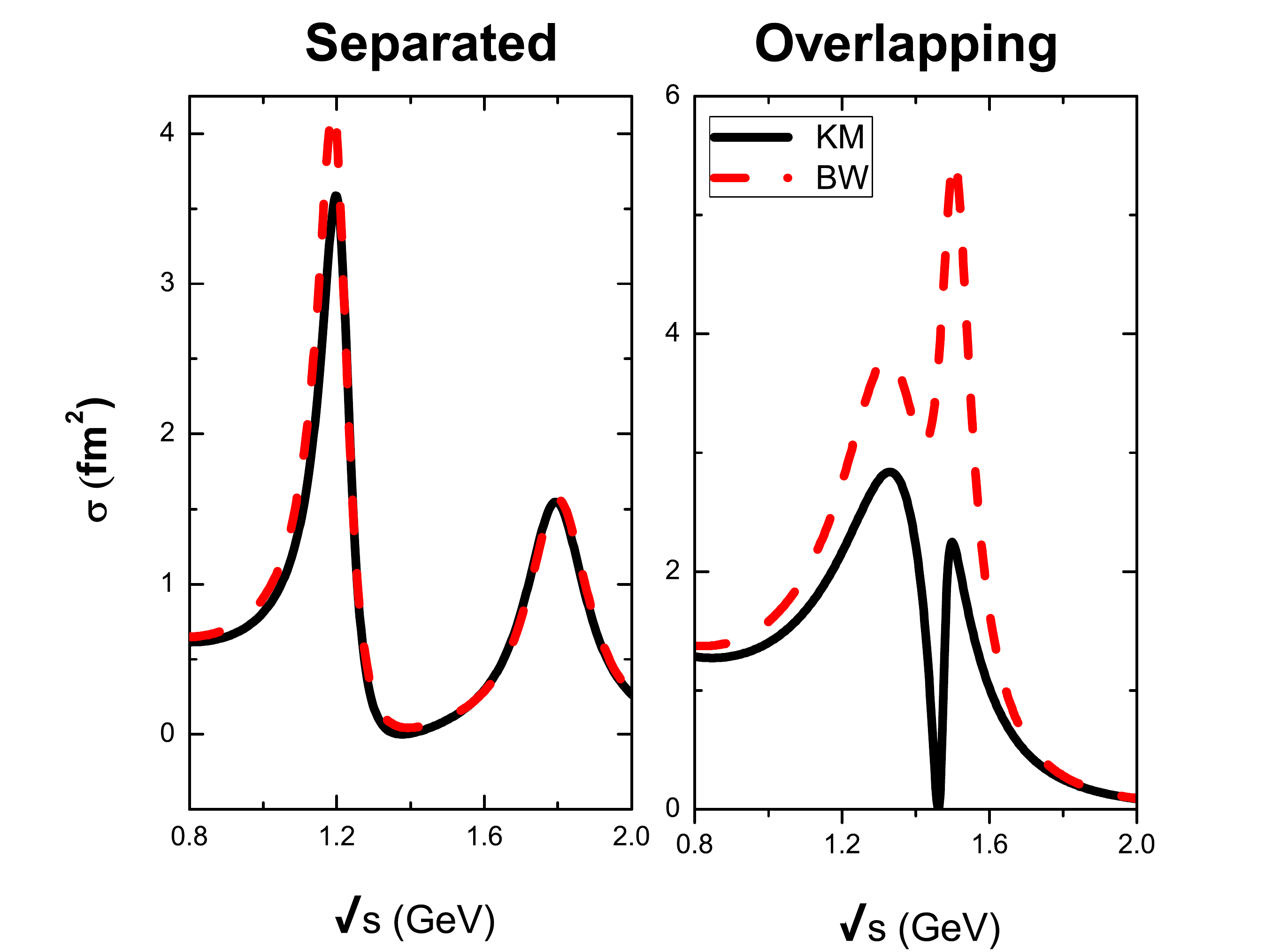} 
\caption{Left panel: Total cross sections for separated resonances $m_\alpha$(1200)
and $m_\beta$ (1800). The  widths are $\Gamma_\alpha = 0.1$ GeV and   $\Gamma_\beta = 0.2$ GeV. Right  panel: Total cross sections for nearby resonances 
$m_\alpha$(1300) with $\Gamma_\alpha=0.3$ GeV and $m_\beta$(1500) with $\Gamma_\beta=0.1$ GeV. 
The solid line is calculation from the $K-$matrix (KM) parameterization and 
the dashed line is calculation from the Breit-Wigner (BW) parameterization.}
\label{tworesonances}
\end{figure}

 In the case of two resonances, such as $a + b \rightarrow (m_\alpha,m_\beta) \rightarrow a + b$,
the $K$-matrix can be written as \cite{Chung:95}
\begin{eqnarray}
 K = \frac{m_\alpha\Gamma_{\alpha\rightarrow \pi\pi}(\sqrt{s})}{m_\alpha^2-s} 
 + \frac{m_\beta\Gamma_{\beta\rightarrow \pi\pi}(\sqrt{s})}{m_\beta^2-s}\,.
\end{eqnarray}
One can then calculate the $T$-matrix from Eq.~(\ref{tmatrixink}) and
the total cross section from  Eq.~(\ref{diffcross}). One can also calculate the total cross section from the BW
parameterization from Eq.~(\ref{bwcross}). To illustrate the scenario when
the two resonances are well separated 
(i.e., when the wings do not significantly overlap), 
we consider the first resonance with
the mass of $m_\alpha  = 1.2$ GeV with a width $\Gamma_\alpha  = 0.1 $ GeV and the second
resonance, $m_\beta = 1.5 $ GeV with a width $\Gamma_\beta = 0.2 $ GeV.
The total cross section for scattering of particles via  these two resonances is shown in the left panel of 
Fig. \ref{tworesonances}. In this case, cross sections from the KM and BW parametrizations are very similar for most 
colliding energies
but the magnitude of the total cross section from the BW parametrization at the peak of the first resonance is slightly larger
than that from the KM parametrization. This is because in BW, the tail of the second resonance contributes to the peak of the 
first resonance which is not the case for the KM. 

\begin{figure}
\includegraphics[width=9cm]{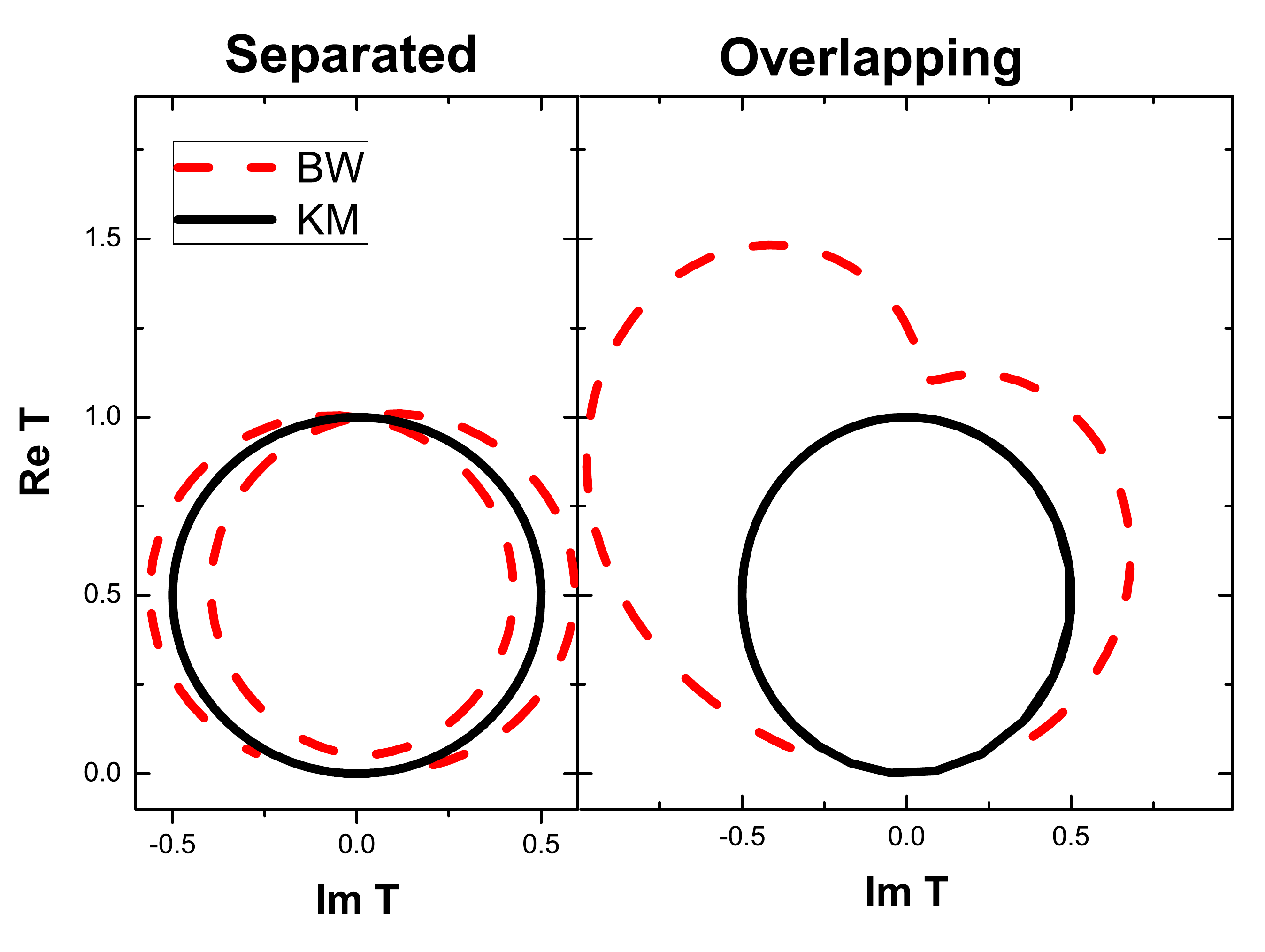}
\caption{Argand diagrams of the T-matrices corresponding to the resonances 
in the left and right panels of Fig. \ref{tworesonances}. 
The solid line is calculation from the $K-$matrix (KM) parameterization and 
the dashed line is calculation from Breit-Wigner (BW) parameterization.}
\label{argandplot}
\end{figure}

We show the Argand plot of the $T$-matrices for both parametrizations in the left panel of Fig. \ref{argandplot}. 
The $T-$matrix from the KM is able to maintain its unitarity whereas that from the BW is unable preserve 
this property exactly.

For two overlapping resonances, we consider  
$m_\alpha = 1.35 $ GeV with $\Gamma_\alpha  = 0.3 $ GeV, and
$m_\beta = 1.5$ GeV and $\Gamma_\beta  = 0.1$ GeV. 
The total cross sections for the KM and BW cases are shown in the right panel of Fig. \ref{tworesonances}.
We can see that the total cross section from the BW parametrization is generally larger than that
from the KM parametrization. The difference is most significant around the peaks of the two
resonances and also in the region between. Such a difference is caused by the contribution from the tail of one resonance
to the peak region of another in the BW parametrization. Such contributions are absent in the KM
parametrization.

In the right panel of Fig. \ref{argandplot}, we show the $T-$matrices from both the KM and BW parameterizations.
The unitarity of the $T-$matrix is preserved in the case of the KM formalism whereas such is not the case in the
BW parameterization.

To conclude this section, we have shown that (i) for interaction through a single resonance, both KM and BW 
parameterization can maintain the unitarity of their $T-$matrices and the total cross sections are very similar 
to that from the PS parameterization, (ii) when an interaction contains multiple resonances,
only the KM formalism is able to maintain the unitarity of the $T-$matrix, and (iii) in the case of multiple resonances, 
the total cross section from the BW is always larger than that from the KM parameterization. 
In this work, we employ the
KM parametrization to calculate the differential cross sections needed for the calculation of viscosity and the phase shifts required for the calculation of entropy density.   Resonance masses and widths are taken from the Particle Data Group (PDG) \cite{PDG:12}. Although the number of instances in which interference effects are significant is small, the use of the KM parametrization automatically guarantees the preservation of unitarity.

\section{Thermodynamics of Interacting Hadrons}

The equilibrium thermal properties of a hadron gas consists of ideal (from translational degrees of motion due to
thermal agitation) and interacting (from strong interaction dynamics) parts. The ideal gas contributions are 
straightforward to compute (see, for example, from the formulas in Appendix 1 of Ref.~\cite{Raju:92}). 
Leading contributions to the thermal properties from interactions between hadrons can be 
calculated using the second virial coefficient which can be deduced from  two-body phase 
shifts \cite{Huang:1963}.  The relativistic  virial expansion introduced by 
Dashen et al \cite{Dashen:69} has been fruitfully employed to calculate 
the state variables of interacting hadrons in Refs. \cite{Welke:90,Raju:92}, 
which we adopt in the following. Explicitly, interacting contributions to the pressure, energy density and 
entropy density at the second virial level are given by the following expressions:
\begin{eqnarray}
n_{int} &=& \frac{1}{\pi^3}\,\int_{M}^\infty d\epsilon\,\epsilon^2K_1(\epsilon\,\beta)\sum_{l,I}{'} g_l^I \delta_l^I(\epsilon)\,,
\label{interactingdensity}\\
 P_{int} &=& \frac{1}{\beta}\,\frac{1}{2\pi^3}\,\int_{M}^\infty d\epsilon\,\epsilon^2\,K_1(\epsilon \beta)
 \sum_{l,I}{'} g_l^I \delta_l^I(\epsilon)\,,
 \label{interactingpressure}\\
 \mathcal{E}_{int} &=& \frac{1}{4\pi^3}\int_{M}^\infty d\epsilon\,\epsilon^3 \left[ K_2(\epsilon \beta)
 + K_0(\epsilon \beta)\right]\sum_{l,I}{'} g_l^I \delta_l^I(\epsilon)\,, \nonumber \\
 \label{interactingenergy} \\
 s_{int} &=& \beta\frac{1}{2\pi^3}\,\int_{M}^\infty d\epsilon\,\epsilon^3\,K_2(\epsilon \beta)
 \sum_{l,I}{'} g_l^I \delta_l^I(\epsilon)\,,
 \label{interactingentropy}
\end{eqnarray}
where $M=m_a+m_b$ is the invariant mass of the interacting pair at threshold, $\beta = 1/T$, $K_\nu$ is the modified Bessel function of order $\nu$, and 
$\epsilon =  (q^2 + m_a^2)^{1/2}  + (q^2 + m_b^2)^{1/2}$ is the total center of mass energy, $q$ being the center of mass momentum of one of the two outgoing particles. The
prime denotes that for a given $l$, the sum over $I$ is resticted to values consistent with statistics $(I+l)$
is even; $g_l^I = (2l+1)(2I+1)$ is the spin-isospin degeneracy of the $(l,I)-$resonance.

It is worthwhile mentioning here that contributions from higher than the second virial coefficient can significantly contribute 
to the state variables mentioned above. In general, the logarithm of the partition function can be written as
\begin{equation}
ln~Z = ln~Z_0 + \sum_{i_1,i_2}  z_1^{i_1}z_2^{i_2}b(i_1,i_2)\, 
\end{equation}
where $z_j=\exp(\beta\mu_j)$ for $j=1,2$ are the fugacities and the virial coefficients $b(i_1,i_2)$ are \cite{Dashen:69}
\begin{eqnarray}
b(i_1,i_2) &=& \frac {V}{4\pi i} \int \frac {d^3P}{(2\pi)^3} \int d\epsilon \exp \left(- \beta (P^2+\epsilon^2)^{1/2} \right) \nonumber \\
&\times& \left[ A \left\{ S^{-1}\frac{\partial S}{\partial \epsilon} - \frac {\partial S^{-1}}{\partial \epsilon} S\right\} \right]_c \,.
\end{eqnarray}
In the above, $V,~P$ and $\epsilon$ stand for the volume, the total center of mass momentum and energy, respectively. The labels $i_1$ and $i_2$ refer to a channel of the $S-$matrix which has an initial state containing $i_1+i_2$ particles --
the trace is therefore over all combinations of particle number. The symbol $A$ denotes symmetrization (anti-symmetrization) operator for a system of bosons (fermions) and the subscript $c$ refers to the trace of all linked diagrams.  The lowest (second) virial coefficient $b_2\equiv b(i_i,i_2)/V$ as $V \rightarrow\infty$ corresponds to the case in which $i_1=i_2=1$ from which the expressions in Eqs. (\ref{interactingdensity}) -- (\ref{interactingentropy}) ensue.   
To our knowledge, multi-particle initial states 
- as would be appropriate, for example, to treat the $\omega(783)$ meson appropriately - have not been attempted thus far (including this work) in calculations of thermodynamics. The work of Lu and Moore \cite{Lu:11}, performed in the context of bulk viscosity,  and in which the $T-$matrix for three pions in the initial state was proposed,  
could be a starting point for investigating the role of the third virial coefficient.

An adequate treatment of repulsive channels and inelastic channels also requires careful attention.    For all but the  lightest mesons ($\pi$ and $K$) and nucleons, experimental knowledge about repulsive interactions are not available.  Since the interacting contributions to the state variables are given by a convolution of the isospin-wieghted sum of the phase shifts with thermal weights, the repulsive channels negate the positive contributions of some of the attractive channels \cite{Raju:92}.
Furthermore, several reactions are characterized by significant amounts of inelasticities.  While a complete knowledge of the 
$T-$ matrix would include inelastic channels, a practical approach that encompasses all the resonances encountered in 
a heavy-ion collision is yet to be devised. In this work, however, we will limit ourselves 
to the level of the second virial coefficient. Improvements to address the limitations mentioned above  will be taken up in a separate work.

%
\begin{figure}
\includegraphics[width=9.5cm]{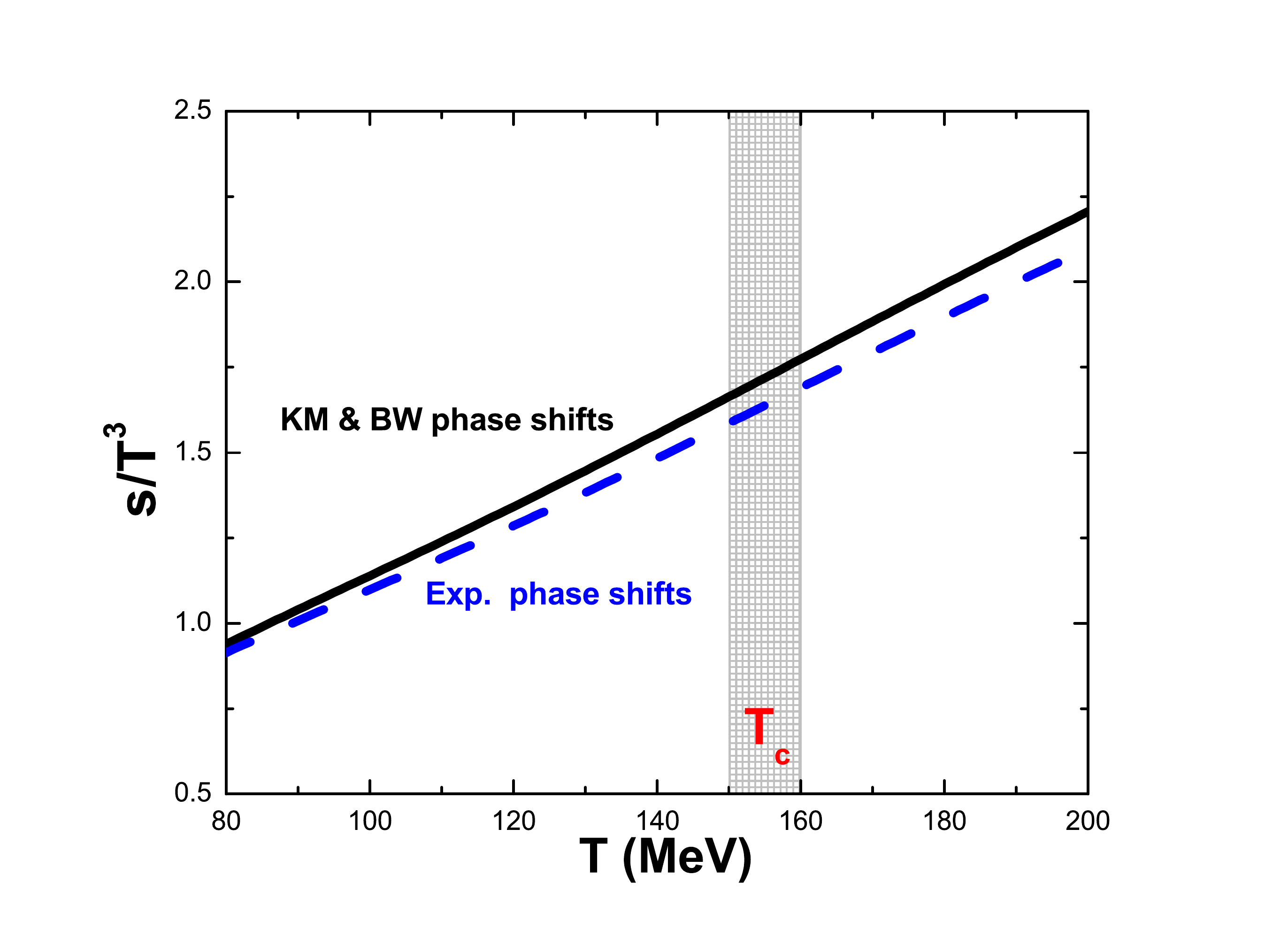}
\caption{Entropy density versus temperature of pions interacting through  the channel $\pi\pi \rightarrow \rho \rightarrow \pi \pi$.
The dashed curve shows results when phase shifts from experiments are 
parameterized as in Eq. (\ref{rhophaseshift}). The solid curve is for the
$K-$matrix (KM) and Breit-Wigner (BW) parametrizations from
Eq. (\ref{phaseandTandK}). The transition temperature, $T_c$, from
Ref. \cite{BMW:12},  is indicated by the  rectangular box
}
\label{entropycompare}
\end{figure}

In the case of $\pi\,\pi \rightarrow \rho \rightarrow \pi\,\pi$, the contribution of interactions, up to the second virial level,  to the entropy density
can be calculated from Eq. (\ref{interactingentropy}) using phase shifts from the KM and BW parametrizations 
or phase shifts fitted to experiments where available.  
In Fig. \ref{entropycompare}, we contrast results of  the total entropy density calculated by adding the ideal part 
to that from 
Eq. (\ref{interactingentropy}) using these different parametrizations of phase shifts.
Again, differences arise from the parametrization of the $\rho-$ resonance in each method. 
It is important to mention here that the KM/BW approaches are able to take account of resonant interactions only from 
information on masses and widths available in the Particle Data book, but they lack consideration of repulsive channels
(as is known to exist in $\pi\pi$ interactions) in many cases. In this and all subsequent figures, the hadron to quark-gluon 
phase transition temperature $T_c=155\pm 5$ MeV from the lattice calculations of the Budapest-Wuppertal collaboration \cite{BMW:12} is shown by a rectangular box. 
All of our results for thermal and transport properties are meaningful only below $T_c$ and the results extending beyond
$T_c$ are only indicative of those in a hypothetical hadronic world.

\section{Single Component System}

Here we consider the case of a system consisting of pions only as a prototype of a single component gas. 
The more realistic case of a multi-component mixture will be considered in  the next section.
In Table \ref{pionchannel}, we show 8 resonant channels in $\pi \pi$ interactions that lead to $\pi \pi$ final states as listed in the Particle Data book \cite{PDG:12}. Those with very small branching ratios have been omitted. 
With increasing temperatures, even this hypothetical single component system is not as simple as one may first assume insofar as many resonances 
can and do contribute to the thermal and transport properties.     

\begin{table}[h]
\begin{tabular}{|cccccccc|}
\hline
Particle  & &  Mass & & Width  & &  & Branching Ratio \\
 & & (MeV) & & (MeV) & & & $\pi-\pi$ \\
\hline
 $\rho$     &  & 774  & & 150 &&  & 1    \\
 $\omega$   &  & 782  & & 8   &&  & 0.02  \\
 $f_0$      &  & 980  & & 100 &&  & 0.7  \\
  $f_2$     &  & 1270 & & 185 &&  & 0.875  \\
 $f_{02}$   &  & 1370 & & 200 &&  & 0.1 \\
 $\rho_2$   &  & 1465 & & 310 &&  & 0.5 \\
  $f_2^{'}$ &  & 1525 & & 76  &&  & 0.01\\
 $\rho_3$   &  & 1690 & & 235 &&  & 0.1 \\
\hline
\end{tabular}
\caption{\label{pionchannel}List of resonances formed  in $\pi-\pi$ interactions.
The first column is the resonance's identity. Resonance masses (second column) and widths (third column) are in units of MeV. The last column gives the branching ratio of each decay channel. 
Resonances with smaller branching ratios than shown are omitted.
Entries are taken from the PDG \cite{PDG:12}.}
\end{table}

\begin{figure}
\includegraphics[width=9.5cm]{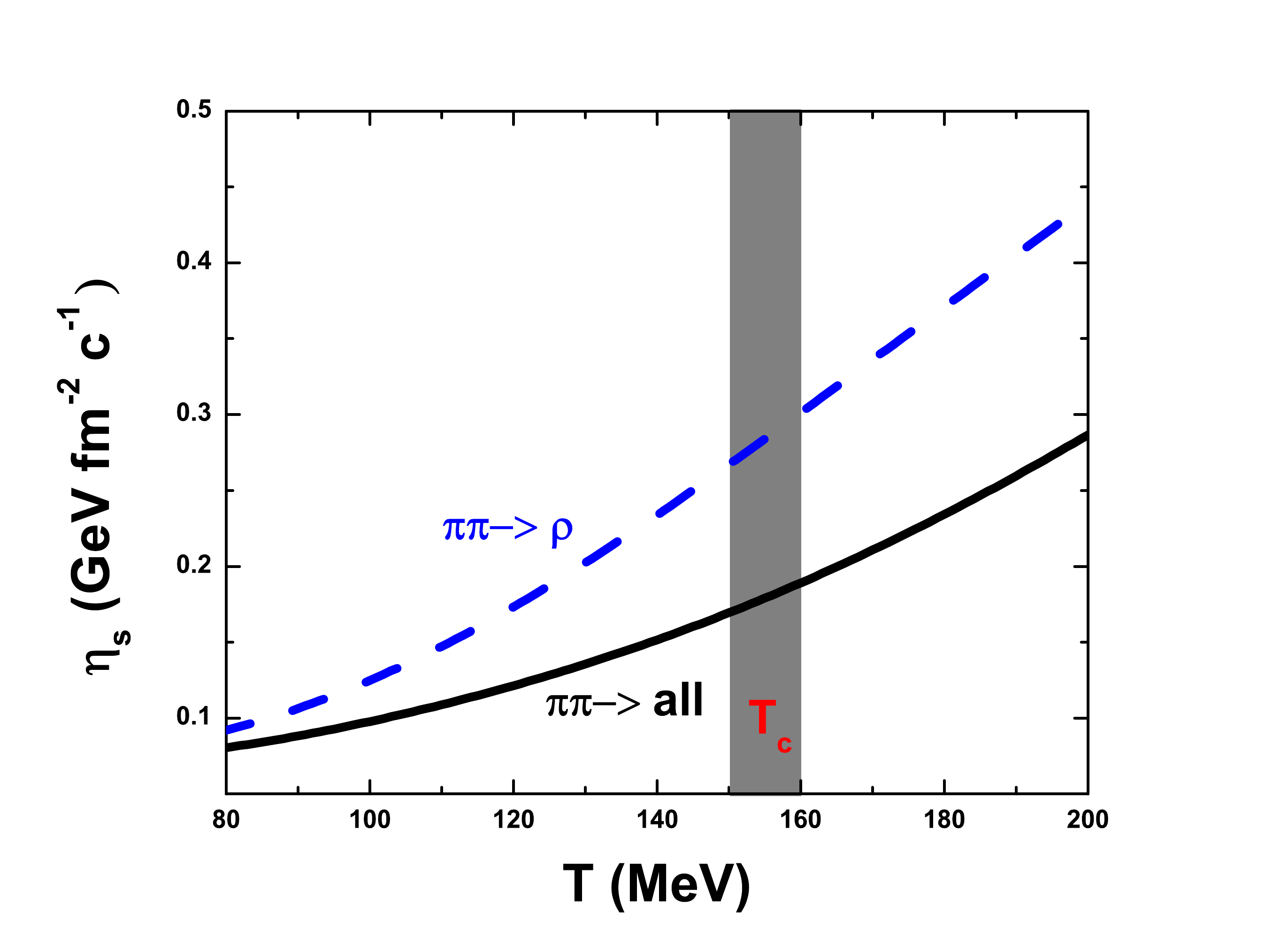}
\caption{Shear viscosities (calculated by using K-Matrix formalism) versus temperature. The dashed curve shows results when
only the $\rho-$resonance was considered. Results when all possible resonances (see Table \ref{pionchannel}) formed in
$\pi-\pi$ interactions were included are shown by the solid curve. 
The transition temperature, $T_c$, is indicated 
by the rectangular box \cite{BMW:12}.}
\label{shearpionpion2}
\end{figure}

The coefficient of shear viscosity for a single component system 
is calculated from Eqs. \ref{shearone}, \ref{c00} and
\ref{relomega} for two different cases: 
(i) $\pi\pi \rightarrow \rho \rightarrow \pi\pi  $ 
and (ii) $\pi\pi \rightarrow \textnormal{All Channels} \rightarrow \pi\pi  $. 
The differential cross-sections for both cases are parameterized by the $K-$matrix formalism described in Sec. \ref{Kmatrixsection} .
The ensuing results for the shear viscosity are shown in Fig. \ref{shearpionpion2}. Although the $\rho-$meson contribution dominates,
the reduction in the magnitude of the viscosity upon the inclusion of all the resonances in Table \ref{pionchannel} is clearly evident, 
particularly close to the phase transition temperature.   

\begin{figure}
\includegraphics[width=9.5cm]{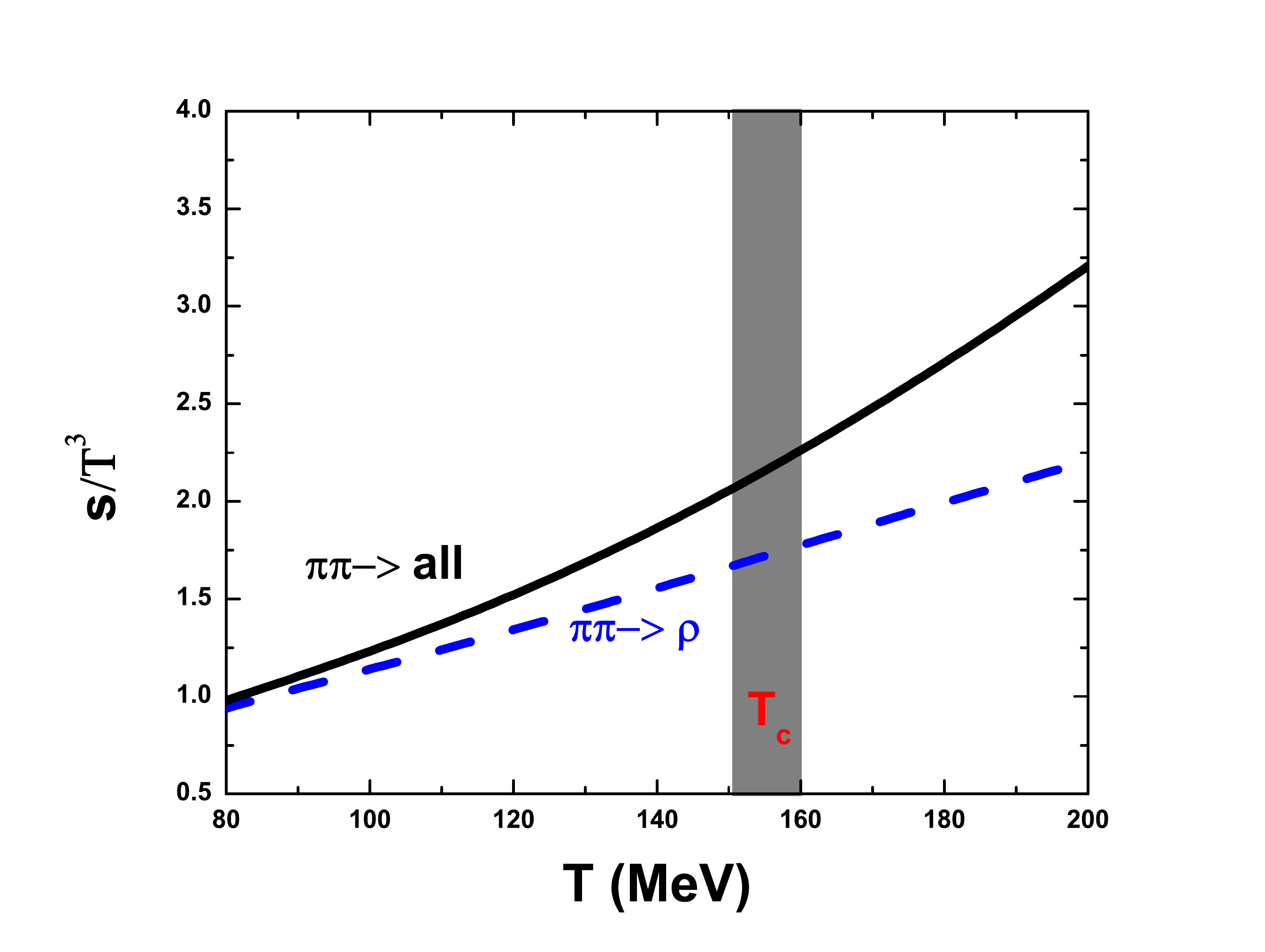}
\caption{Entropy density versus temperature of an interacting pion gas. The dashed line is the result when only
the $\rho-$channel is included whereas the solid line is 
the result when all 
resonances in Table \ref{pionchannel} are included. 
The transition temperature, $T_c$, is indicated 
by the  rectangular box \cite{BMW:12}.}
\label{spiongas}
\end{figure}

The addition of more channels increases the magnitudes of 
thermodynamics quantities, such as the pressure, energy density and entropy density. 
In Fig. \ref{spiongas}, we show the entropy density which is calculated by adding the ideal gas contributions to those from 
Eq. (\ref{interactingentropy}) for the $\pi\pi$ interactions. 
Relative to the case when only the $\rho-$ resonance is considered (dashed curve), 
the significant increase in the entropy density when all resonances in  Table \ref{pionchannel} are included 
(solid curve)
is chiefly due to the increased number of (spin and isospin) degrees of freedom.

\begin{figure}
\includegraphics[width=9.5cm]{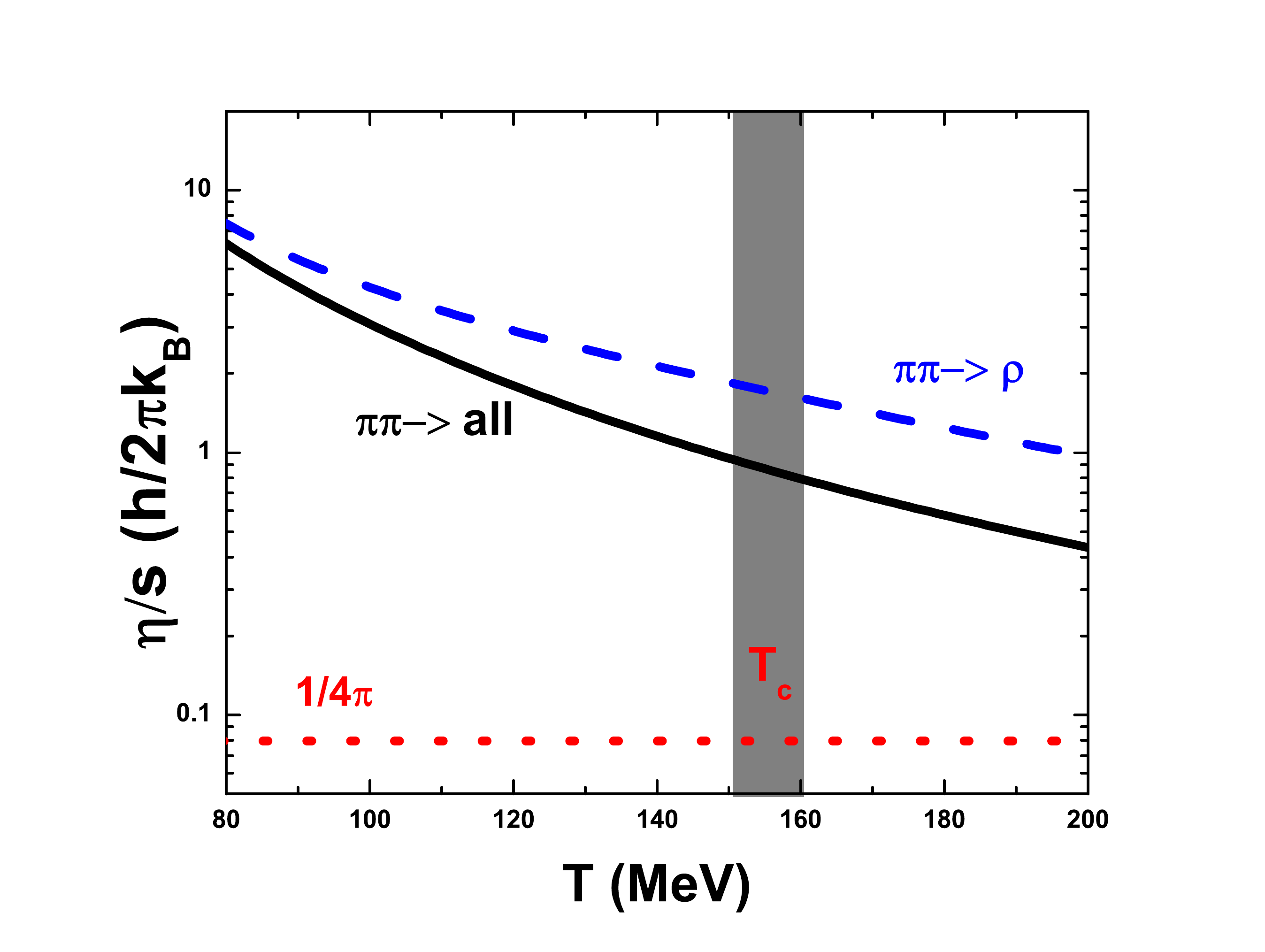}
\caption{Ratio of shear viscosity (calculated using the K-Matrix formalism) to entropy density versus temperature. The dashed line corresponds
to the case when only the $\rho-$resonance was considered. The solid line shows results when all possible resonances formed in
$\pi-\pi$ interactions (see Table \ref{pionchannel}) were included. The AdS/CFT lower bound of $\eta/s$ is shown by the red dashed-line. 
The transition temperature, $T_c$, is indicated 
by the  rectangular box \cite{BMW:12}.}
\label{etaspion}

\end{figure}

The ratios of $\eta/s$ for the cases discussed above are shown in Fig. \ref{etaspion}. 
The dashed line is the result when only the $\rho-$channel is considered and 
the solid line shows the result when all resonances  in Table. \ref{pionchannel} are included. 
As the temperature approaches 
the hadron to quark-gluon phase transition temperature 
$T_c =  155 \pm 5$ MeV found in lattice simulations \cite{BMW:12}, $\eta/s$ decreases more rapidly
when all channels are included than when only the $\rho-$ channel is considered. 
The inclusion of several resonances not only causes a reduction in 
the coefficient of shear viscosity, $\eta$,  it also results in a significant  
increase in the entropy density, $s$. Both of these effects render 
the ratio $\eta/s$ to become small as the temperature approaches $T_c$.

\section{Multi-Component system}
In this section, we consider a four component mixture comprising of $\pi-K-N-\eta(548)$ and include 
the dominant resonances (see Tables  \ref{pionchannel} and \ref{allresonances}) produced in binary elastic interactions among the various constituents. 
In order to gain an understanding of how the inclusion of increasing number of particles (and their 
associated resonances) determines the magnitudes of $\eta$ and $\eta/s$, we also show results 
for the two-component mixtures of  $\pi-K$ and $\pi-N$, 
as well as for the three-component mixture of $\pi-K-N$ together with results for the $\pi-K-N-\eta(548)$ mixture.
Thermodynamic properties for each of the above systems, calculated using the virial expansion formalism of Sec. IV, 
are presented in terms of the interaction measure $I=(\epsilon-3P)/T^4$  and compared 
with the lattice results of the Budapest-Wuppertal (BW) collaboration \cite{BMW:12} as well as with those of a hadron resonance gas (HRG) with the same 
constituents from Ref. \cite{Jan:13}.
\begin{table*}\centering
\begin{tabular}{@{}|rrrrrrrrrrrrcr|@{}}\toprule
\hline
& &  & &  & & \multicolumn{5}{c}{Branching Ratios} & \phantom{abcde}& &\\
\cmidrule{3-6}
Particle  & &  Mass  & & Width   & &  &  $\pi-K$ & $K-K$ & $\pi-\eta$ &$\pi-N$ & $K-N$ & &\\
\hline
 $K^{*}$ &   & 893&  & 50 &&  &   1. & - & - & -  & -  & & \\
 $K_0^{*}$ &   & 1429&   & 287 &&   &  1. & - & -  & -  & -  &  &\\
 $K_2$ &   &  1430&  & 100 &&  &   0.5 & - & - & -  & -  & & \\
 $K_2^{*}$ &   & 1410&   &227 &&   &  0.3 & - & -& -  & -  & & \\
 $K_3^{*}$  &  & 1680&   &323 &&   &  0.4 & - & - & -  & -  & & \\
  \hline
  $a_0$ &   & 984&  & 185 &&   &  - &0.1  & 0.9 & -  & -  &  & \\
 $\phi$ &   & 1020&  & 4 &&  &  - & 0.84 & - & -  & -   &  & \\
  $f_0^{*}$ &   & 1370&   & 200 &&   &  - & 0.2 & -  & -   & -  & & \\
 $a_2$ &   & 1320&  &  107 &&  &   - & 0.05 & 0.14 & -   & -  &  & \\
 $\phi_2$ &   & 1680 &   & 150 &&   &  - & 0.1 & - & -  & -  &  & \\
 $\phi_3$ &   & 1890 &   & 400 &&   &  - & 0.1 & - & -   & -  &   & \\
 \hline
  $\Delta_{1232}$ &   & 1232&   & 115 &&   &  - & -   &  -   &1. & -  & & \\
  $\Delta_{1600}$ &   & 1600&   & 200 &&   &  - & -   &  -   &0.15 & -  &  & \\
  $\Delta_{1620}$  &  & 1620&   & 180 &&   &  - & -   &  -   &0.25 & -  &  & \\
  $\Delta_{1700}$  &  & 1700&   & 300 &&   &  - & -   &  -   &0.20 & -  &  & \\
  $\Delta_{1900}$  &  & 1900&   & 240 &&   &  - & -   &  -   &0.30 & -  &  & \\
  $\Delta_{1905}$  &  & 1905&   & 280 &&   &  - & -   &  -   &0.20 & -  &  & \\
  $\Delta_{1910}$  & &  1910&   & 280 &&   &  - & -   &  -   &0.3 & -  &  & \\
  $\Delta_{1920}$  &  & 1920&   & 260 &&   &  - & -   &  -   &0.15 & -  &  & \\
  $\Delta_{1930}$  &  & 1930&   & 360 &&   &  - & -   &  -   &0.15 & -  &  & \\
  $\Delta_{1950}$  &  & 1950&   & 285 &&   &  - & -   &  -   &0.45 & -  &  & \\
  \hline
  $N^{*}_{1440}$  &  & 1440&   & 200 &&   &  - & -   &  -  &0.7 & -  &  & \\
  $N^{*}_{1520}$  &  & 1520&   & 125 &&   &  - & -   &  -  &0.6 & -  &  & \\
  $N^{*}_{1535}$  &  & 1535&   & 150 &&   &  - & -   &  -  &0.55 & -  &  & \\
  $N^{*}_{1650}$  &  & 1650&   & 150 &&   &  - & -   &  -  &0.8 & -  &  & \\
  $N^{*}_{1675}$  &  & 1675&   & 140 &&   &  - & -   &  -  &0.45 & -  &  & \\
  $N^{*}_{1680}$  &  & 1680&   & 120 &&   &  - & -   &  -  &0.65 & -  &  & \\
  $N^{*}_{1700}$  &  & 1700&   & 100 &&   &  - & -   &  -  &0.10 & -  &  & \\
  $N^{*}_{1710}$  &  & 1710&   & 110 &&   &  - & -   &  -  &0.15 & -  &  & \\
  $N^{*}_{1720}$  &  & 1720&   & 150 &&   &  - & -   &  -  &0.15 & -  &  & \\
  $N^{*}_{1900}$  &  & 1900&   & 250 &&   &  - & -   &  -  &0.1 & -  &  & \\
  $N^{*}_{1990}$  &  & 1990&   & 550 &&   &  - & -   &  -  &0.05 & -  &  & \\
  \hline
  $\Lambda^{*}_{1520}$  &  & 1520&   & 16 &&   &  - & -   &  -  & -  &0.45 &  & \\
  $\Lambda^{*}_{1600}$  &  & 1600&   & 150 &&   & - & -   &  -  & -  &0.35 &  & \\
  $\Lambda^{*}_{1670}$  &  & 1670&   & 35 &&   &  - & -   &  -  & -  &0.20 &  & \\
  $\Lambda^{*}_{1690}$  &  & 1690&   & 60 &&   &  - & -   &  -  & -  &0.25 &  & \\
  $\Lambda^{*}_{1800}$  &  & 1800&   & 300 &&   & - & -   &  -  & -  &0.40 &  & \\
  $\Lambda^{*}_{1810}$  &  & 1810&   & 150 &&   & - & -   &  -  & -  &0.35 &  & \\
  $\Lambda^{*}_{1820}$  &  & 1820&   & 80 &&   &  - & -   &  -  & -  &0.65 &  & \\
  $\Lambda^{*}_{1830}$  &  & 1830&   & 95 &&   &  - & -   &  -  & -  &0.10 &  & \\
  $\Lambda^{*}_{1890}$  &  & 1890&   & 100 &&   & - & -   &  -  & -  &0.35 &  & \\
  \hline
  $\Sigma^{*}_{1660}$  &  & 1660&   & 100 &&   &  - & -   &  -  & - &0.3 &  & \\
  $\Sigma^{*}_{1670}$  &  & 1670&   & 60 &&   &  - & -   &  -  & -  &0.13 &  & \\
  $\Sigma^{*}_{1750}$  &  & 1750&   & 90 &&   &  - & -   &  -  & -  &0.4 &  & \\
  $\Sigma^{*}_{1775}$  &  & 1775&   & 120 &&   & - & -   &  -  & -  &0.4 &  & \\
  $\Sigma^{*}_{1915}$  &  & 1915&   & 120 &&   & - & -   &  -  & -  &0.15 &  & \\
  $\Sigma^{*}_{1940}$  &  & 1940&   & 220 &&   & - & -   &  -  & -  &0.1&  & \\
  \hline
\bottomrule
\end{tabular}
\caption{List of Resonances involved in $\pi-K-\eta-N$ interactions.
The first column contains the resonance's identity. Masses (second column) and widths (third column)  are in units of MeV. 
The remaining columns give  
the branching ratios of the corresponding decay channels. The data shown are from the PDG, Ref.  \cite{PDG:12}.}
\label{allresonances}
\end{table*}

\begin{figure}
\includegraphics[width=9.5cm]{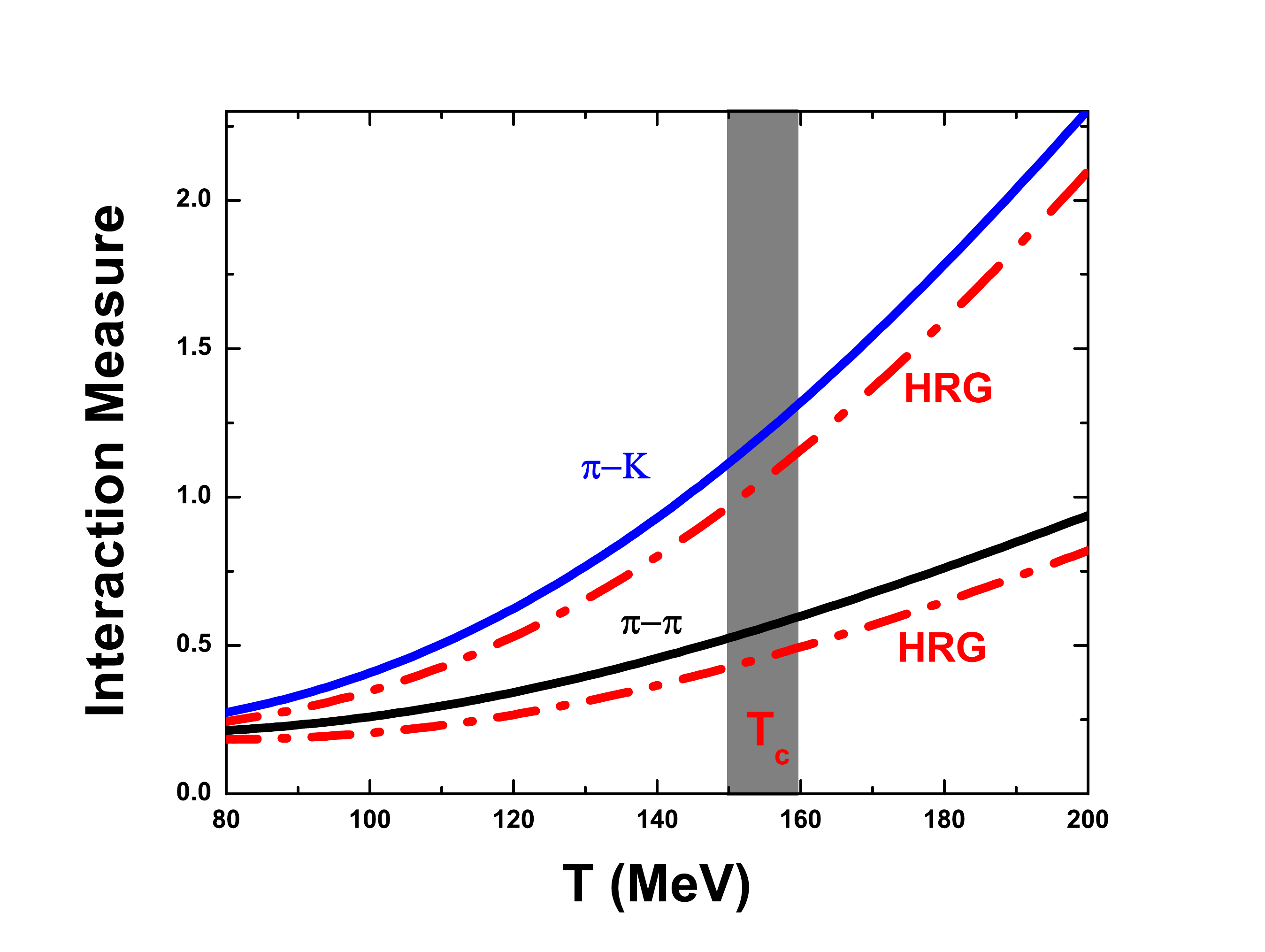}
\caption{Interaction measure $I = (\epsilon - 3P)/T^4$ of interacting pions 
and of the mixture $\pi-K$. The solid lines are results from the
virial expansion approach and the dashed lines are results from a Hadron Resonances Gas (HRG)
model \cite{Jan:13}.
The transition temperature, $T_c$ is indicated 
by the rectangular box \cite{BMW:12}. }
\label{interactionmeasureHRG_no_br}
\end{figure}

In Fig. \ref{interactionmeasureHRG_no_br}, the interaction measure $I$ is shown for the pion gas and for the
binary mixture $\pi-K$.  Our results using the virial expansion method are calculated 
from Eqs. (\ref{interactingpressure}) and (\ref{interactingenergy}) and are shown by the solid
curves. Results of the HRG model \cite{Jan:13}, calculated using the same number of particles/resonances,
are  shown by the dashed curves. 
 For the pion gas, all resonances from Table. \ref{pionchannel} are included
in both calculations. For the $\pi-K$ mixture,  
interactions between $\pi-\pi$, $\pi-K$ and $K-K$ through the 
resonances shown in Tables. \ref{pionchannel} and \ref{allresonances} are included. 
Note that the results from the virial expansion approach are slightly larger compared
to those from the HRG model. We attribute this  
difference to the fact that the widths of the various resonances are accounted for naturally 
in the former approach whereas the HRG model includes resonances without consideration of 
their widths.  In the limit of sufficiently small widths, the virial expansion approach guarantees that
the pressure (and other thermodynamic quantities) of a dilute system of interacting particles is 
very well approximated by that of an ideal gas of noninteracting particles including the resonances 
(see Refs. \cite{Dashen:69,Dashen:74,Rajaraman:79,Raju:92}).
This feature forms the basis of the HRG model.   However, repulsive channels (as, for example, the $\delta_0^2$ channel in 
$\pi\pi$ interactions) are not included in the HRG model. Where available, effects of repulsive channels can be straightforwardly 
incorporated in the virial expansion approach through the appropriate phase shifts 
(unfortunately, adequate data does not exist for all but the lightest mesons).

\begin{figure}
\includegraphics[width=9.5cm]{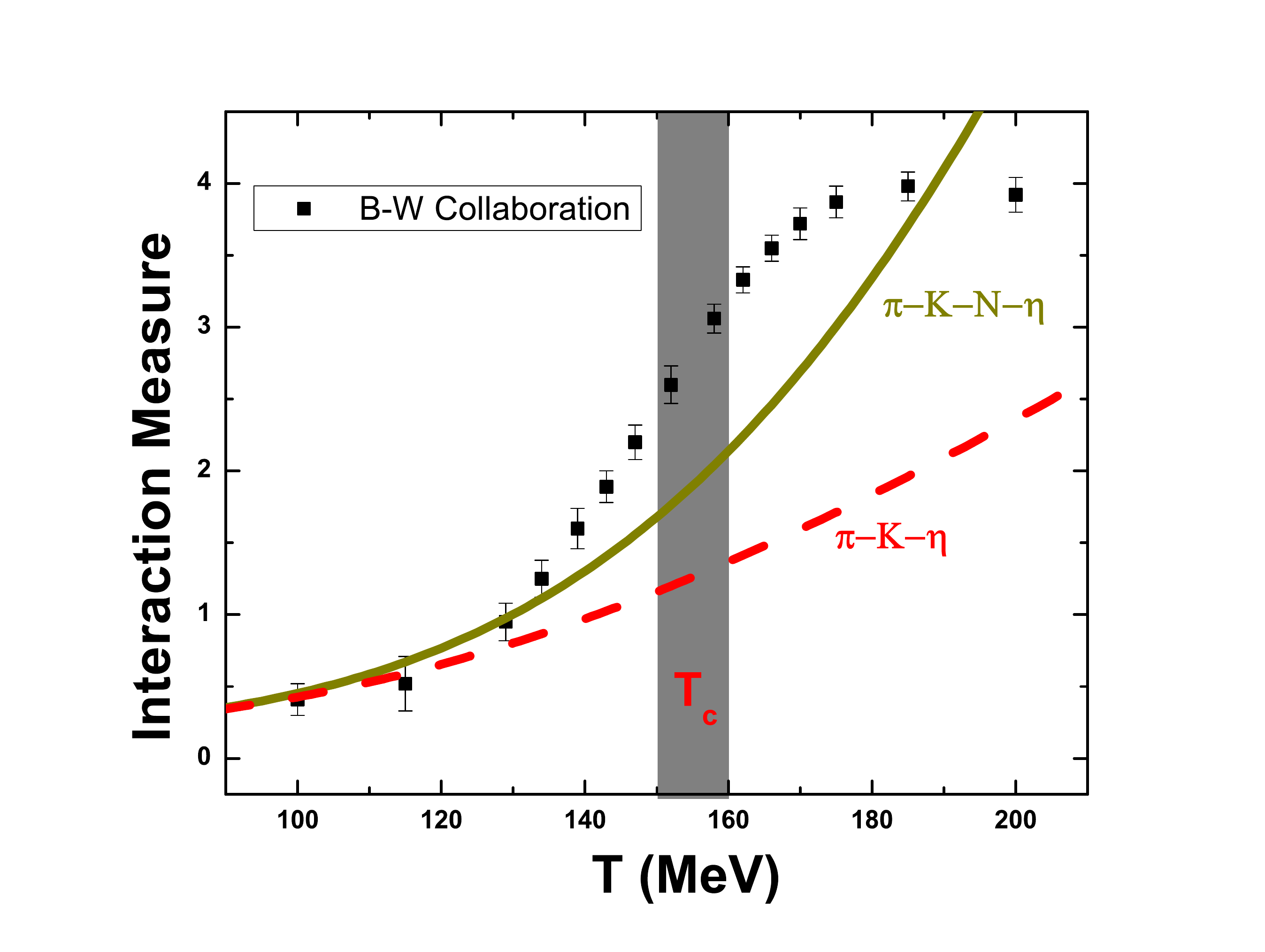}
\caption{Interaction measure $I = (\epsilon - 3P)/T^4$ 
from lattice calculations (data points) with 3-quarks ($n_f = 2 +1$),  
$m_\pi = 135$ MeV and lattice spacing, $N_t = 8$ (BMW Collaboration).
The solid line is the result for a three component mixture of $\pi-K-\eta(548)$
and the dashed-dotted line is the result for a  four component mixture of $\pi-K-\eta(548)- N$.
The transition temperature, $T_c$ is indicated 
by the rectangular box \cite{BMW:12}.}
\label{interactionmeasure_no_br}
\end{figure}

A comparison of results from the virial expansion approach to those from lattice calculations 
of the BW collaboration \cite{BMW:12} is shown in Fig. \ref{interactionmeasure_no_br}.
In the interest of being consistent with
the ingredients of transport calculations, only those resonances that are formed 
in the $\pi-K-\eta-N$ mixture (shown in Tables \ref{pionchannel} and \ref{allresonances}) are included in the virial expansion approach.  
The number of resonances included in the case of the  $\pi-K-\eta$ mixture is 21 whereas 57 resonances
are included in the  $\pi-K-\eta-N$ mixture. 
As is evident from this figure, the additional resonances present in the four component
mixture improve the agreement with the lattice results up to 140 MeV.  The inclusion of 
additional mesons and baryons more massive than realized in 
the $\pi-K-\eta-N$ mixture is expected to improve agreement with the current lattice results (as 
supported by results of HRG calculations with all the resonances in the PDG book)
even up to the phase transition temperature $T_c$ \citep{BMW:12}.

\begin{figure}
\includegraphics[width=9.5cm]{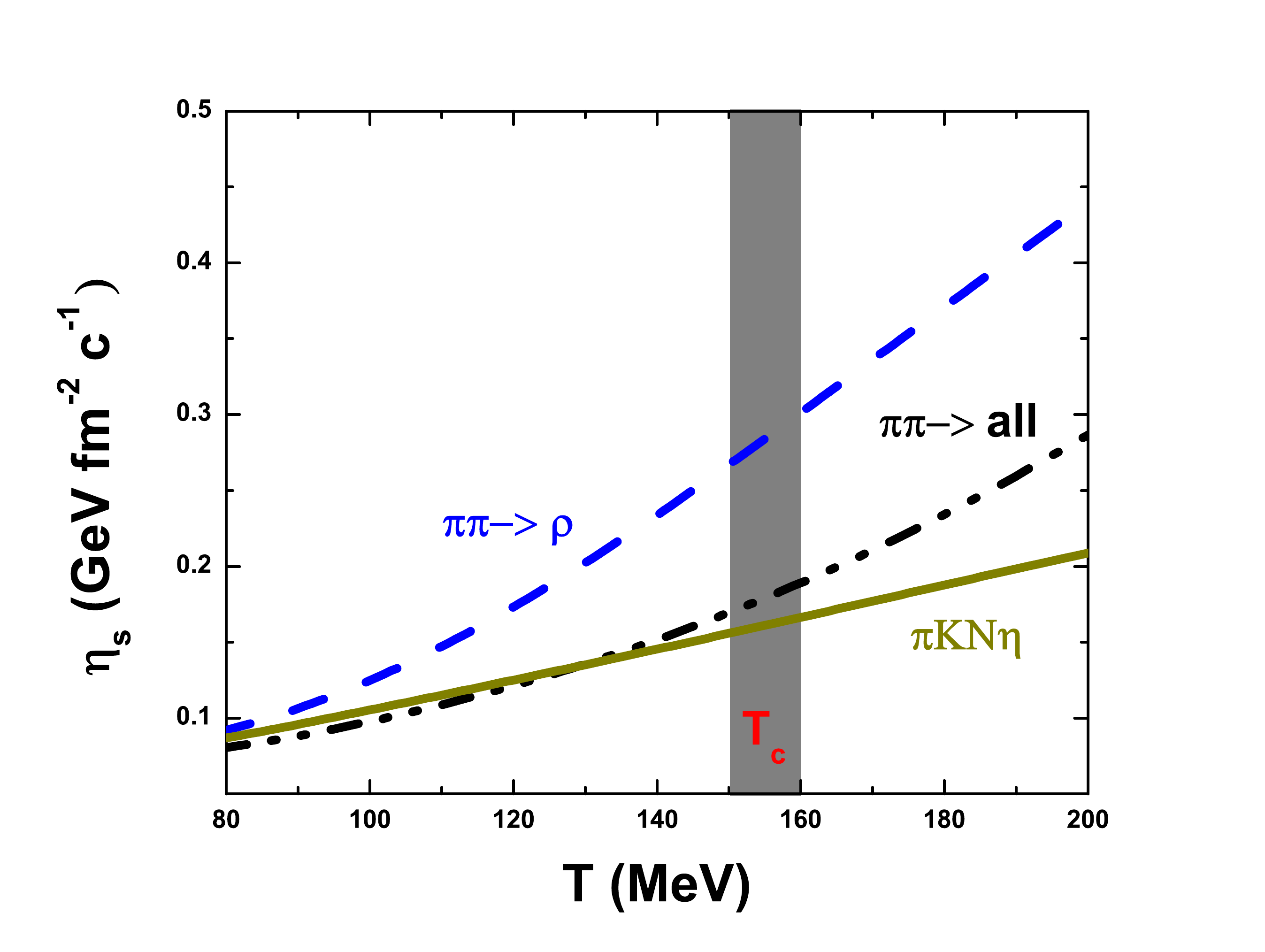}
\caption{Shear viscosity versus temperature for the $\pi-K-\eta-N$ mixture.
The transition temperature, $T_c$ is indicated by the rectangular box
\cite{BMW:12}.}
\label{etapion4comp}
\end{figure}

The temperature dependence of the shear viscosity in a $\pi-K-\eta-N$ mixture 
is displayed in Fig. \ref{etapion4comp}. The progressive decrease in the 
magnitude of the shear viscosity with increasing temperature as more and more resonances
are included is readily apparent from this figure particularly as $T_c$ is approached. 
The inclusion of more resonances than considered in this work 
is likely to reduce the magnitude of $\eta$ even further as $T_c$ is approached.

\begin{figure}
\includegraphics[width=9.5cm]{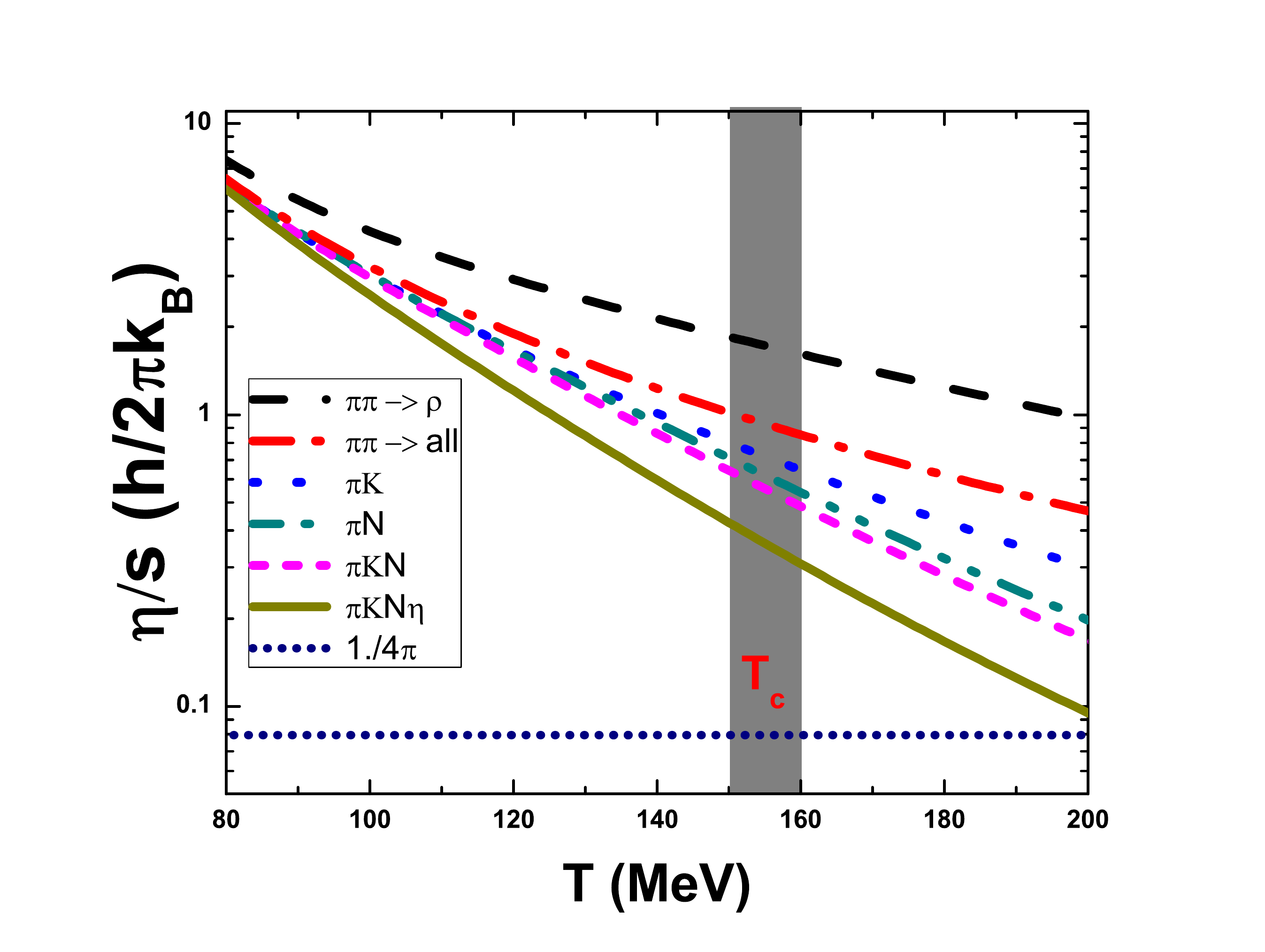}
\caption{Ratio of shear viscosity to entropy density, $\eta/s$, of 
interacting hadrons. Results are for a single component system ($\pi$ gas), two component mixtures
($\pi-K$ and $\pi-N$), a three component mixture ($\pi-K-N$) and 
a four component mixture ($\pi-K-\eta(548)-N$). 
The horizontal curve at $1/(4\pi)$ is the AdS/CFT result.}
\label{PKNETA4mixture}
\end{figure}

The results of the ratio $\eta/s$ are presented in Fig. \ref{PKNETA4mixture}. 
As explained in the 
previous section, the role of increasing the number of resonances is evident even in the case of a single component 
pion gas.  In addition to decreasing the shear viscosity, resonances increase the entropy density both effects serving 
to decrease $\eta/s$ with increasing temperature. A similar trend is observed in the case binary mixtures as seen 
from the results for the $\pi-K$ and $\pi-N$ mixtures. The fact that $\eta/s$ for these two mixtures are nearly the same 
is intriguing. A physical understanding of this result resides in the masses of the resonances realized in these two systems 
which are nearly the same.  
Results for the three ($\pi-K-N$) and four ($\pi-K-\eta(548)-N$) component systems, 
highlight the increasing role of the enhanced entropy density in these systems as the heaviest resonances
are not very effective in transferring momentum in a direction perpendicular to that of fluid flow.
For reference, the AdS/CFT result of $1/(4\pi)$ is also shown in this figure. From the trends seen in these results,
we infer that the inclusion of additional mesons and baryons will further decrease $\eta/s$ as $T_c$ is approached though
not to the level of the AdS/CFT result. 

\section{Summary and Conclusion }

In this paper, we have calculated the shear viscosity $\eta$ and the entropy density $s$ of an interacting hadronic resonance gas.
Calculations of $\eta$ were performed using the relativistic generalization of the Chapman-Enskog approach, whereas calculations of
the entropy density employed the relativistic virial expansion method. Our results are for a  hadronic mixture with net baryon number zero and 
are meaningful only below the hadron to quark-gluon phase transition temperature of about 155 MeV.
The mixture considered comprised of four basic hadrons $\pi-K-\eta-N$ in which 57 resonances, formed through binary 
elastic interactions between the  constituents, were included. In order to understand the results for the four-component mixture, calculations for a single 
component pion gas, binary mixtures
$\pi-K$ and $\pi-N$, and  the tertiary mixture $\pi-K-\eta$ with dominant resonances
were also performed. In both the calculations of $\eta$ and $s$, phase shifts 
for binary interactions among the various particles feature prominently. Experimental information on such phase shifts are only available for 
a few cases in the $\pi-\pi$, $\pi-K$, $\pi-N$ and $N-N$ systems, but not for most of the massive resonances formed.  From the masses and widths
of these various resonances, phase shifts (necessary in the calculation of the thermodynamic properties of the system using the virial expansion
approach) and associated differential cross sections (required in the calculation of shear viscosity using the Chapman-Enskog method)
were obtained using both the Breit-Wigner (BW) and the K-matrix (KM) formalisms. For narrow and well separated resonances, both of these formalisms 
yield similar results that maintain unitarity of the T-matrix. However, for wide and nearby resonances only the K-matrix approach preserves unitarity. The use of phase shifts 
and differential cross sections obtained from the K-matrix approach in the calculations of $\eta$ and $s$ is the principal new element 
of this work.       

Our results for the single component pion gas in which 8 resonances up to 1700 MeV are included 
highlights the important role of resonances in determining the magnitudes of $\eta$ and $s$ as a 
function of temperature. Increasing 
number of resonances 
decreases the magnitude of $\eta$ as additional channels of interaction are opened up, but increases $s$ due  
to the increase in the number of degrees of freedom. Both these features render the ratio $\eta/s$ small.
These physical effects are further fortified in binary mixtures
of $\pi-K$ and $\pi-N$ due to the presence of a large number of resonances 
resulting in a further decrease of the ratio $\eta/s$ from the case of the single component 
pion gas. Results for the tertiary mixture containing $\pi-K-\eta$ and the four component mixture $\pi-K-N-\eta$
continue the trends observed in the fewer component mixtures both for $\eta$ and $s$ resulting in a much reduced ratio of
$\eta/s$.  The implication of these results is that were more mesons and baryons than considered in this work to be
included, the ratio $\eta/s$ would become even smaller than what we find here  for the  $\pi-K-N-\eta$ mixture, though
perhaps not to the level of the AdS/CFT result of $1/(4\pi)$ as $T_c$ is approached. 

A few future directions are suggested by our work. The inclusion of more mesons and baryons than considered here 
will reveal how low $\eta/s$ can become in a system of hadrons with zero net baryon and strangeness numbers. To be of practical use for the low-energy  
scans of RHIC and other upcoming accelerators, calculations of both $\eta$ and $s$ for finite baryon and strangeness numbers appear
worthwhile. These tasks will be taken up in separate works.

\section*{Acknowledgements}

This work is supported by the Natural Science Foundation of China under grant No. 11221504.
Research support from US Department of Energy under contract  \#DE-AC02-05CH11231(for AW, VK and XNW)
and Central China Normal University through colleges of
basic research and operation of MOE (for AW and XNW) is gratefully acknowledged.
MP acknowledges research support from the U.S. Department of Energy under contract \#DE-FG02-93ER-40756. 

\appendix

\section{Phase Shift and Breit-Wigner Cross Sections}
Here, we discuss the extent to which cross sections derived from phase shifts 
and the Breit-Wigner parametrized are related. To be specific, 
we consider the process $\pi \pi \rightarrow \rho \rightarrow \pi \pi$ and its associated 
experimental phase shift which is parametrized as  
\begin{eqnarray}
 \delta_1^1(\sqrt{s}) = \frac{\pi}{2} + 
 \arctan\left(\frac{\sqrt{s}- m_\rho}{\Gamma_{\rho\rightarrow 2\pi}(\sqrt{s})/2} \right)\,.
\end{eqnarray}
The corresponding differential cross section is
\begin{eqnarray}
 \sigma(s,\theta) = \frac{C(I,l)}{q^2}\,\sin^2\delta_1^1\, P_1(\cos \theta) ,
\end{eqnarray}
where $C(I,l)$ is the symmetry factor which contains spin-isospin multiplicities. 
We can recast the above expression 
to resemble the well-known Breit-Wigner formula using trigonometric identities: 
\begin{eqnarray}
 \sin(a+b) = \sin(a)\cos(b) + \cos(a)\sin(b)\,,
\end{eqnarray}
where $a = \pi/2$ and $b = \arctan((\sqrt{s}-m_\rho)/(\Gamma_{\rho\rightarrow 2\pi}/2))$, and 
\begin{eqnarray}
  \cos^2(b) = \frac{1}{1 + \tan^2(b)} \,,
\end{eqnarray}
whence 
\begin{eqnarray}
 \sin^2\delta_1^1 = \frac{1}{1 + 
 \frac{(\sqrt{s}-m_\rho)^2}{\Gamma_{\rho\rightarrow 2\pi}^2/4}} = 
 \frac{\Gamma_{\rho\rightarrow 2\pi}^2/4}{(\sqrt{s}-m_\rho)^2 + \Gamma_{\rho\rightarrow 2\pi}^2/4}\nonumber\\
\end{eqnarray}
so that  the differential cross section can be written as 
\begin{eqnarray}
 \sigma(s,\theta) = \frac{C(I,l)}{q^2}\frac{\Gamma_{\rho\rightarrow 2\pi}^2/4}{(\sqrt{s}-m_\rho)^2 
 + \Gamma_{\rho\rightarrow 2\pi}^2/4}
 P_1(\cos \theta)\nonumber\\
\end{eqnarray}
which is the Breit-Wigner formula. However, some differences exist in the parametrizations of the widths:  
\begin{eqnarray}
 \Gamma_{\rho\rightarrow 2\pi}(\textnormal{PS}) &=& 0.095\,q\,\left(\frac{q/m_\pi}
 {1 + (q/m_\rho)^2} \right)^2 \\
 \Gamma_{\rho\rightarrow 2\pi}(\textnormal{BW}) &=& \Gamma_0 \left(\frac{q}{\sqrt{s}} \right)
 \left(\frac{m_\rho}{q_\rho} \right)\left[B^1(q,q_\rho) \right]^2\,.
\end{eqnarray}
Numerical values of $\Gamma$'s and their total cross sections
at three representative energies, $\sqrt{s} = 2m_\pi, m_\rho$ and $ \infty$
shown in Table. \ref{gammacomparison} serve to illustrate the differences. 
\begin{table}[h]
\caption{\label{gammacomparison}Widths of the $\rho-$resonance at three energies.}
\begin{ruledtabular}
\begin{tabular}{|c|c|c|c|c|c|}
$\sqrt{s} $ &  $q/\sqrt{s}$ & $\Gamma_{\rho\rightarrow 2\pi}(\textnormal{PS})$
& $\Gamma_{\rho\rightarrow 2\pi}(\textnormal{BW})$ & $\sigma_\textnormal{PS}(s)$ & $\sigma_\textnormal{BW}(s)$\\
\hline
 $2m_\pi$ &  0 & 0 & 0 & 0 & 0\\
 $m_\rho$ &   $q_\rho/m_\rho$ & $\Gamma_0 = 0.155$ & $\Gamma_0 = 0.155$ & $C(I,l)/q_\rho^2$ &
 $C(I,l)/q_\rho^2$\\
 $\infty$ &   $1/2$ & 0 & 0.252 & 0 & 0\\
\end{tabular}
\end{ruledtabular}
\end{table}

\section{The $K-$Matrix and Breit-Wigner Cross Sections}
The squared $T-$matrix can be written as 
\begin{eqnarray}
 |T|^2 = \frac{1}{(1+K^2)^2} \left(K^2 +K^4 \right) = \frac{K^2}{(1+K^2)}\,.
\end{eqnarray}
In the case of the $\rho-$resonance, one has
\begin{eqnarray}
K^2 &=& \frac{m_\rho^2\Gamma_{\rho\rightarrow 2\pi}^2}{(m_\rho^2 - s)^2} \\
(1+K^2)^2 &=& \frac{(m_\rho^2 - s)^2 + m_\rho^2\Gamma_{\rho\rightarrow 2\pi}^2}{(m_\rho^2-s)^2} 
\end{eqnarray}
which allows us to write 
\begin{eqnarray}
 |T|^2 &=& \frac{m_\rho^2\Gamma_{\rho\rightarrow 2\pi}^2}{(m_\rho^2-s)^2 + m_\rho^2\Gamma_{\rho\rightarrow 2\pi}^2}\,.
\end{eqnarray}
Near the peak of the resonance, 
\begin{eqnarray}
 m_\rho^2 - s = (m_\rho - \sqrt{s})(m_\rho + \sqrt{s})\simeq(m_\rho -\sqrt{s})2m_\rho \nonumber \\
\end{eqnarray}
so that 
\begin{eqnarray}
 |T|^2 = \frac{\Gamma_{\rho\rightarrow 2\pi}^2/4}{(m_\rho-\sqrt{s})^2 + \Gamma_{\rho\rightarrow 2\pi}^2/4}\,.
\end{eqnarray}
Note that near the peak of the resonance, the cross sections obtained
from phase shifts, the Breit-Wigner formula, and the $K-$matrix
formalism are all the same.
If the same width is used in all three parametrizations, the resulting cross sections
will be identical.

\section{$K-$matrix Cross Section for a Single Resonance}
The differential cross section for a binary interaction is given by
\begin{eqnarray}
\sigma(s,\theta)_{ab\rightarrow cd} &=& |f_{ab\rightarrow cd}(s,\theta)|^2 \,,
\end{eqnarray}
where $f_{ab\rightarrow cd}(s,\theta)$ is the scattering amplitude given by
\begin{eqnarray}
f_{ab\rightarrow cd}(s,\theta) &=& \frac{1}{q} \sum_l (2l+1) T^l_{ab\rightarrow cd}(s) P_l(\cos \theta)\,,
\end{eqnarray}
with $ q = 0.5\,\sqrt{s - 4\,m_\pi^2}$
, $l$ is the orbital angular momentum, $T^l_{ab\rightarrow cd}(s)$ is the 
interaction matrix and $P_l(\cos \theta)$ is the orbital angular 
momentum dependence of the $T$-matrix. In terms of the $K-$ matrix, 
the $T-$matrix can be obtained from 
\begin{eqnarray}
 \textnormal{Re}\, T = \frac{K}{(1+K^2)}\,\, \textnormal{and}\,\,
 \textnormal{Im}\, T = \frac{K^2}{(1+K^2)}\, .
\end{eqnarray}
For a resonant interaction (using the example of 
the $\rho-$ resonance to be specific), the $K-$matrix can be written as 
\begin{eqnarray}
 K &=& \frac{m_\rho \Gamma_{\rho\rightarrow 2\pi}}{m_\rho^2 - s}\,
\end{eqnarray}
where $\Gamma_{\rho\rightarrow 2\pi}$ is the width of the $\rho-$resonance: 
\begin{eqnarray}
 \Gamma_{\rho\rightarrow 2\pi} &=& \Gamma_{\rho\rightarrow 2\pi}^0\,
 \left(\frac{m_\rho}{\sqrt{s}} \right) \left(\frac{q}{q_\rho} \right)
 \left[B^1(q,q_\rho) \right]^2\,,
\end{eqnarray}
where $\Gamma_{\rho\rightarrow 2\pi}^0$ is the width at the pole, 
$q_\rho = 0.5\,(m_\rho^2 - 4\,m_\pi^2)^{1/2}$ and the factor 
$B$ is given by 
\begin{eqnarray}
 B^1(q,q_\rho) &=& \frac{F_1(q)}{F_1(q_\rho)}\,,
\end{eqnarray}%

where
\begin{eqnarray}
 F_1(q) = \sqrt{\frac{2\,z}{z+1}}\,\,,\,\,\,
 z(q) =\frac{q}{0.197} \,.
\end{eqnarray}

\section{Two and More Resonances}
When two or more resonances are involved, the path through the $T-$matrix within the $K-$matrix formalism becomes cumbersome, albeit straightforward.
An alternative way is to use the scattering amplitude. Recall that
\begin{eqnarray}
 \sigma(s,\theta) = \left| f_{ab \rightarrow cd}(s,\theta) \right|^2 \,,
\end{eqnarray}
the scattering amplitude being 
\begin{eqnarray}
 f_{ab \rightarrow cd}(s,\theta) = \sum_l^\infty \left( 2l + 1 \right) f_l(\sqrt{s}) P_l(\cos \theta)
\end{eqnarray}
and
\begin{eqnarray}
 f_l(\sqrt{s}) = \frac{e^{i\delta_l} \sin\,\delta_l }{q} = 
 \frac{1}{q} \left( \cos \delta_l \sin \delta_l + i \sin^2\delta_l \right) \,.
\end{eqnarray}
In the case of a single resonance, 
\begin{eqnarray}
 \left| f_{ab \rightarrow cd}(s,\theta) \right|^2 = (2l+1)^2\frac{\sin^2 \delta_l}{q^2} P_l^2(\cos \theta) \,,
\end{eqnarray}
while in the case of two resonances, 
\begin{eqnarray}
 f_1 &=& \frac{1}{q} \left( \cos \delta_1 \sin \delta_1 + i \sin^2 \delta_1 \right) P_1(\cos \theta) \\
 f_2 &=& \frac{1}{q} \left( \cos \delta_2 \sin \delta_2 + i \sin^2 \delta_2 \right) P_2(\cos \theta) \,.
\end{eqnarray}
Resolving the scattering amplitude into its real and imaginary parts,  we obtain 
\begin{eqnarray}
 \textnormal{Re}\,f_{ab \rightarrow cd} &=& \frac{1}{q} \left((2l_1+1) \cos \delta_1 \sin \delta_1 P_1(\cos \theta) \right. \nonumber \\
 && \left. +\, (2l_2+1) \cos \delta_2 \sin \delta_2 P_2(\cos \theta) \right)  \\
 \textnormal{Im}\,f_{ab \rightarrow cd} &=& \frac{1}{q} \left( (2l_1+1) \sin^2\delta_1 P_1(\cos \theta)  \right. \nonumber \\
 && \left. + \, (2l_2+1) \sin^2\delta_2 P_2(\cos \theta) \right)\,.
\end{eqnarray}
The generalization to more more than two resonances proceeds along similar lines as above.

%
%


\bibliographystyle{h-physrev3}%
\bibliography{kmref}

\end{document}